\documentclass[12pt]{article}
\usepackage{a4wide}
\usepackage{latexsym}
\usepackage{amsmath}
\usepackage{amsfonts}
\usepackage{amscd}
\usepackage{cite}

\usepackage{pslatex}
\usepackage{graphicx}
\usepackage[latin1,utf8]{inputenc}
\usepackage[T1]{fontenc}

\allowdisplaybreaks

\newcommand{\bq}{\begin{eqnarray}}
\newcommand{\eq}{\end{eqnarray}}
\newcommand{\eps}{\varepsilon}
\newcommand{\Eulerconstant}{\gamma_{\mathrm{EM}}}

\newcommand{\iterint}[1]{F\left(#1\right)}

\usepackage{color}

\begin{document}

\thispagestyle{empty}

\begin{flushright}
  MITP/19-046 
\end{flushright}

\vspace{1.5cm}

\begin{center}
  {\Large\bf 
   The unequal mass sunrise integral expressed through iterated integrals on $\overline{\mathcal M}_{1,3}$ \\
  }
  \vspace{1cm}
  {\large Christian Bogner ${}^{a}$, Stefan M\"uller-Stach ${}^{b}$ and Stefan Weinzierl ${}^{a}$ \\
  \vspace{1cm}
      {\small ${}^{a}$ \em Institut f{\"u}r Physik,}\\
      {\small \em Johannes Gutenberg-Universit{\"a}t Mainz,}\\
      {\small \em D - 55099 Mainz, Germany}\\
  \vspace{2mm}
      {\small ${}^{b}$ \em Institut f{\"u}r Mathematik,}\\
      {\small \em Johannes Gutenberg-Universit{\"a}t Mainz,}\\
      {\small \em D - 55099 Mainz, Germany}\\
  } 
\end{center}

\vspace{2cm}

\begin{abstract}\noindent
  {
We solve the two-loop sunrise integral with unequal masses systematically to all orders in the dimensional
regularisation parameter $\varepsilon$.
In order to do so, we transform the system of differential equations for the master integrals to an $\varepsilon$-form.
The sunrise integral with unequal masses depends on three kinematical variables.
We perform a change of variables to standard coordinates on the moduli space ${\mathcal M}_{1,3}$
of a genus one Riemann surface with three marked points.
This gives us the solution as iterated integrals on $\overline{\mathcal M}_{1,3}$.
On the hypersurface $\tau=\mbox{const}$ our result reduces to elliptic polylogarithms.
In the equal mass case our result reduces to iterated integrals of modular forms.
   }
\end{abstract}

\vspace*{\fill}

\newpage

\section{Introduction}
\label{sec:intro}

The theory of Feynman integrals, which evaluate to multiple polylogarithms is by now very well understood.
But not all Feynman integrals may be expressed in terms of multiple polylogarithms.
Already at two-loops one encounters in quantum field theory Feynman integrals, which are associated to elliptic curves
and cannot be expressed in terms of 
multiple polylogarithms 
\cite{Laporta:2004rb,MullerStach:2011ru,Adams:2013nia,Bloch:2013tra,Remiddi:2013joa,Adams:2014vja,Adams:2015gva,Adams:2015ydq,Bloch:2016izu,Adams:2017ejb,Bogner:2017vim,Adams:2018yfj,Honemann:2018mrb,Bloch:2014qca,Sogaard:2014jla,Tancredi:2015pta,Primo:2016ebd,Remiddi:2016gno,Adams:2016xah,Bonciani:2016qxi,vonManteuffel:2017hms,Adams:2017tga,Ablinger:2017bjx,Primo:2017ipr,Passarino:2017EPJC,Remiddi:2017har,Bourjaily:2017bsb,Hidding:2017jkk,Broedel:2017kkb,Broedel:2017siw,Broedel:2018iwv,Lee:2017qql,Lee:2018ojn,Adams:2018bsn,Adams:2018kez,Broedel:2018qkq,Bourjaily:2018yfy,Bourjaily:2018aeq,Besier:2018jen,Mastrolia:2018uzb,Ablinger:2018zwz,Frellesvig:2019kgj,Broedel:2019hyg,Blumlein:2019svg,2019arXiv190611857B}.
Similar integrals occur in string theory \cite{Broedel:2014vla,Broedel:2015hia,Broedel:2017jdo,DHoker:2015wxz,Hohenegger:2017kqy,Broedel:2018izr}.

The method of differential equations \cite{Kotikov:1990kg,Kotikov:1991pm,Remiddi:1997ny,Gehrmann:1999as,Argeri:2007up,MullerStach:2012mp,Henn:2013pwa,Henn:2014qga,Ablinger:2015tua,Adams:2017tga,Bosma:2017hrk}
combined with integration-by-parts identities \cite{Tkachov:1981wb,Chetyrkin:1981qh,Laporta:2001dd}
is widely used to compute Feynman integrals.
One first expresses all relevant Feynman integrals in terms of few master integrals $\vec{J}$, sets up a system of differential equations for the latter
and solves this system of differential equations with appropriate boundary conditions.
The solution is particularly simple, if the system of differential equations can be brought into an
$\eps$-form \cite{Henn:2013pwa}:
\bq
\label{eps_form}
 d \vec{J}
 & = &
 \eps A \vec{J},
\eq
where the matrix $A$ does not depend on the dimensional regularisation parameter $\eps$.
In order to achieve this form, one seeks a suitable basis of master integrals $\vec{J}$, related to a pre-canonical basis $\vec{I}$ by
\bq
 \vec{J} & = & U \vec{I}.
\eq
The entries of the matrix $A$ are differential one-forms.
In addition we would like to have that these entries have a standard form. 
For example in the case of multiple polylogarithms we would like to have that they are of the form
\bq
 d \ln\left(p\left(\vec{x}\right)\right),
\eq
where $p(\vec{x})$ is a polynomial in the kinematic variables $\vec{x}$.
In order to achieve this, a change of the kinematic variables may be necessary:
\bq
 \vec{x} & \rightarrow & \vec{x}'
\eq
In the case of multiple polylogarithms the transformation $U$ from a Laporta basis $\vec{I}$ to the basis $\vec{J}$ is algebraic
in the kinematic variables. 
If the transformation is rational in the kinematic variables, several algorithms exist to find such a 
transformation \cite{Gehrmann:2014bfa,Argeri:2014qva,Lee:2014ioa,Prausa:2017ltv,Gituliar:2017vzm,Meyer:2016slj,Meyer:2017joq,Lee:2017oca,Adams:2017tga,Becchetti:2017abb}. 
Already in the case of multiple polylogarithms it might be necessary to perform a change of kinematic variables 
in order to rationalise square roots \cite{Becchetti:2017abb,Lee:2017oca,Besier:2018jen,Chaubey:2019lum,Becchetti:2019tjy}.

Let us now turn to the elliptic case.
The simplest Feynman integral which cannot be expressed in terms of multiple polylogarithms is given by
the two-loop equal mass sunrise integral \cite{Broadhurst:1993mw,Berends:1993ee,Bauberger:1994nk,Bauberger:1994by,Bauberger:1994hx,Caffo:1998du,Laporta:2004rb,Kniehl:2005bc,Groote:2005ay,Groote:2012pa,Bailey:2008ib,MullerStach:2011ru,Adams:2013nia,Bloch:2013tra,Adams:2014vja,Adams:2015gva,Adams:2015ydq,Remiddi:2013joa,Bloch:2016izu,Adams:2017ejb,Bogner:2017vim,Adams:2018yfj,Honemann:2018mrb,Groote:2018rpb}.
The equal mass sunrise integral depends on a single scale $x=p^2/m^2$ and is related to a single elliptic curve.
The system of differential equations for this integral can be brought to an $\eps$-form \cite{Adams:2018yfj}.
In this case, transcendental functions (a period of the elliptic curve) appear in the transformation matrix $U$.
If in addition one changes the kinematic variable from $x$ to the modular parameter $\tau$ \cite{Bloch:2013tra},
one identifies the entries of the matrix $A$ as modular forms \cite{Adams:2017ejb}.
The equal mass sunrise integral can therefore be expressed to all orders in the dimensional regularisation parameter as iterated integrals
of modular forms.
The same applies to several other integrals, like the kite integral, depending on a single scale and related to a single elliptic curve.

We may now ask what happens if a Feynman integral depends on more than one scale.
In this article we consider the sunrise integral with unequal masses as a prototype for a Feynman integral with several scales, but still
related to a single elliptic curve.
The sunrise integral with unequal masses depends on three kinematical variables.
In this paper we show that even for the sunrise integral with unequal masses
there is a transformation of the master integrals, which brings the system of differential equations
into an $\eps$-form as in eq.~(\ref{eps_form}).
In addition we perform a change of the kinematic variables towards standard coordinates for the moduli space of a Riemann surface of genus
one with three marked points.
The integration kernels appearing in the matrix $A$ are then differential forms on $\overline{\mathcal M}_{1,3}$ with logarithmic singularities
only.
The integration kernels involve the functions $g^{(k)}(z,\tau)$, obtained from the expansion of the Kronecker function.
These functions are not double-periodic and hence not single-valued on ${\mathcal M}_{1,3}$.
We view them as multi-valued functions on ${\mathcal M}_{1,3}$.
Alternatively, we may consider instead of ${\mathcal M}_{1,3}$ a covering space of ${\mathcal M}_{1,3}$.
In practice, this does not pose any major problem. Our aim is to integrate a differential equation from a given
boundary point. This is harmless as long as we don't cross a branch point.
In the vicinity of a branch point we just have to ensure that we continue continously to the next sheet.

Our result unifies the result in the equal mass case, expressed in terms of iterated integrals of modular forms, 
with the available partial results for the first few
terms in the $\eps$-expansions in the unequal mass case \cite{Adams:2014vja,Adams:2015gva,Bloch:2016izu,Broedel:2017siw}, expressed as elliptic polylogarithms.
Elliptic polylogarithms are iterated integrals on an elliptic curve with 
fixed modular parameter $\tau$ and several marked points \cite{Beilinson:1994,Levin:1997,Levin:2007,Enriquez:2010,Brown:2011,Wildeshaus}.
We would like to point out that the natural setting for the sunrise integral with unequal masses
is the moduli space $\overline{\mathcal M}_{1,3}$, where the modular parameter $\tau$ is one of the coordinates.

This paper is organised as follows:
In the next section we introduce our notation.
In section~\ref{sec:elliptic_curve} we discuss the elliptic curve(s) associated to the sunrise integral with unequal masses.
We may either associate an elliptic curve obtained from the Feynman graph polynomial or an elliptic curve obtained from the maximal cut.
These two curves are not isomorphic, but isogenic.
In section~\ref{sect:variables} we introduce the kinematic variables which we will use. These are standard coordinates
on ${\mathcal M}_{1,3}$.
In section~\ref{sect:special_functions} we introduce the functions and the one-forms, which will appear in the matrix $A$.
In section~\ref{sect:masters} we present the master integrals, giving us a system of differential equations in $\eps$-form.
This system of differential equations is presented in section~\ref{sect:differential_equation}.
In section~\ref{sect:iterated_integrals} we briefly review iterated integrals as a preparation for
section~\ref{sect:analytical_results}, where we present the analytic results for the master integrals
associated to the sunrise integral with unequal masses.
The results are given as iterated integrals on $\overline{\mathcal M}_{1,3}$.
Finally, our conclusions are given in section~\ref{sect:conclusions}.
In an appendix we summarise our notation for some standard mathematical functions, 
give details on the relation between two isogenic elliptic curves
and describe the content of the supplementary electronic file attached to the arxiv version of this article.

\section{Notation}
\label{sec:notation}

\subsection{The sunrise integral in momentum space}

The sunrise integral with general masses is defined by
\bq
\label{def_momentum_space}
 S_{\nu_1 \nu_2 \nu_3}\left( D, \eps, p^2, m_1^2, m_2^2, m_3^2, \mu^2 \right)
 & = &
 e^{2\Eulerconstant \eps}
 \left(\mu^2\right)^{\nu_{123}-D}
 \int \frac{d^Dk_1}{i \pi^{\frac{D}{2}}} \frac{d^Dk_2}{i \pi^{\frac{D}{2}}}
 \frac{1}{D_1^{\nu_1} D_2^{\nu_2} D_3^{\nu_3}},
 \;\;\;
\eq
where $D$ denotes the number of space-time dimensions,
$\Eulerconstant$ denotes the Euler-Mascheroni constant, 
$\mu$ is an arbitrary scale introduced to render the Feynman integral dimensionless,
and the quantity $\nu_{123}$ is defined by
$\nu_{123}=\nu_1+\nu_2+\nu_3$.
The inverse propagators are given by
\bq
 & &
 D_1=-k_1^2+m_1^2, 
 \hspace{5mm}  
 D_2=-k_2^2+m_2^2, 
 \hspace{5mm}  
 D_3 = -(p-k_1-k_2)^2+m_3^2.
\eq
For $(p^2,m_1^2,m_2^2,m_3^2)=(0,0,0,0)$ the integral is zero.
For $(p^2,m_1^2,m_2^2,m_3^2)=(p^2,0,0,0)$ the integral is given by
\bq
\lefteqn{
 S_{\nu_1 \nu_2 \nu_3}\left( D, \eps, p^2, 0, 0, 0, \mu^2 \right)
 = } & & 
 \nonumber \\
 & &
 e^{2\Eulerconstant \eps}
 \left(\frac{-p^2}{\mu^2}\right)^{D-\nu_{123}}
 \frac{\Gamma\left(\nu_{123}-D\right)}{\Gamma\left(\nu_1\right)\Gamma\left(\nu_2\right)\Gamma\left(\nu_3\right)}
 \frac{\Gamma\left(\frac{D}{2}-\nu_1\right)\Gamma\left(\frac{D}{2}-\nu_2\right)\Gamma\left(\frac{D}{2}-\nu_3\right)}{\Gamma\left(\frac{3D}{2}-\nu_{123}\right)}.
\eq
Excluding these special cases, we may assume that at least one internal mass is non-zero, let this be $m_3$.
Unless stated otherwise we set in the following
\bq
 \mu & = & m_3
\eq
and 
\bq
 D & = & 2 - 2\eps.
\eq
The integral depends then only on $\eps$ and the ratios 
\bq
 x \; = \; \frac{p^2}{m_3^2}, 
 \;\;\;\;\;\;
 y_1 \; = \; \frac{m_1^2}{m_3^2}, 
 \;\;\;\;\;\;
 y_2 \; = \; \frac{m_2^2}{m_3^2}.
\eq
We write with a slight abuse of notation
\bq
 S_{\nu_1 \nu_2 \nu_3}\left( \eps, x, y_1, y_2 \right)
 & = &
 S_{\nu_1 \nu_2 \nu_3}\left( 2-2\eps, \eps, p^2, m_1^2, m_2^2, m_3^2, m_3^2 \right).
\eq
Considering the sunrise integral in $2-2\eps$ dimensions instead of $4-2\eps$ dimensions is no restriction:
With the help of dimensional recurrence relations \cite{Tarasov:1996br,Tarasov:1997kx} 
one recovers the result in $4-2\eps$ dimensions from the result in $2-2\eps$ dimensions.

\subsection{The Feynman parameter representation}

The Feynman parameter representation of the sunrise integral is given by
\bq
 S_{\nu_1 \nu_2 \nu_3}\left( \eps, x, y_1, y_2 \right)
 & = &
 e^{2 \Eulerconstant \eps}
 \frac{\Gamma(\nu_{123}-2+2\eps)}{\Gamma(\nu_1)\Gamma(\nu_2)\Gamma(\nu_3)}
 \int\limits_{\sigma} 
 \left( \prod\limits_{j=1}^{3} \alpha_j^{\nu_j-1} \right)
 \frac{{\mathcal U}^{\nu_{123}-3+3\eps}}{{\mathcal F}^{\nu_{123}-2+2\eps}}
 \omega,
\eq
where the integration is over
\bq
 \sigma & = & \left\{ \left[ \alpha_1 : \alpha_2 : \alpha_3 \right] \in {\mathbb R} {\mathbb P}^2 | \alpha_i \ge 0 \right\}.
\eq
The differential form $\omega$ is given by
\bq
 \omega & = & 
 \alpha_1 \; d\alpha_2 \wedge d\alpha_3
 + \alpha_2 \; d\alpha_3 \wedge d\alpha_1
 + \alpha_3 \; d\alpha_1 \wedge d\alpha_2.
\eq
The graph polynomials are given by
\bq
\label{def_U_and_F}
 {\mathcal U}
 & = & 
 \alpha_1 \alpha_2 + \alpha_2 \alpha_3 + \alpha_3 \alpha_1,
 \nonumber \\
 {\mathcal F} 
 & = & 
 - \alpha_1 \alpha_2 \alpha_3 x 
 + \left( \alpha_1 y_1 + \alpha_2 y_2 + \alpha_3 \right) {\mathcal U} .
\eq
In particular, the integral $S_{111}(0,x,y_1,y_2)$ is given in the Feynman parameter representation by
\bq
 S_{111}\left( 0, x, y_1, y_2 \right)
 & = &
 \int\limits_{\sigma} 
 \frac{\omega}{{\mathcal F}}.
\eq

\subsection{The Baikov representation}

The Baikov representation \cite{Baikov:1996iu,Lee:2009dh,Kosower:2011ty,CaronHuot:2012ab,Frellesvig:2017aai,Bosma:2017ens,Harley:2017qut}
is obtained from the loop momentum representation by 
a change of variables from a subset of the $2 D$ loop momentum variables
to the inverse propagators.
In order to express any scalar product involving the loop momenta, we have to introduce one auxiliary inverse propagator.
Let this be $D_4=-(k_1+k_2)^2$.
The remaining loop momentum variables are integrated out.
The Baikov representation reads
\bq
 S_{\nu_1 \nu_2 \nu_3}\left( \eps, x, y_1, y_2 \right)
 \sim
 \frac{e^{2\Eulerconstant \eps} \left(m_3^2\right)^{\nu_{123}-2+2\eps}}{4 \pi \left(\Gamma\left(\frac{1}{2}-\eps\right)\right)^2}
 \left(-p^2\right)^\eps
 \int dD_1 dD_2 dD_3 dD_4
 \frac{D_4^\eps G_1^{-\frac{1}{2}-\eps} G_2^{-\frac{1}{2}-\eps} }{D_1^{\nu_1} D_2^{\nu_2} D_3^{\nu_3}},
\eq
with the Gram determinants $G_1$ and $G_2$ given by
\bq
 G_1
 & = &
 - \frac{1}{4} \left[ \left( D_1-D_2 \right)^2 + D_4^2 - 2 \left( D_1 + D_2 \right) D_4 - 2 \left(m_1^2-m_2^2\right) \left(D_1-D_2\right) + 2 \left(m_1^2+m_2^2\right) D_4 
 \right. \nonumber \\
 & & \left.
 + \left(m_1^2-m_2^2\right)^2 \right],
 \nonumber \\
 G_2 
 & = &
 - \frac{1}{4} \left[ \left(D_3-D_4\right)^2 + 2 \left(p^2-m_3^2 \right) D_3 + 2 \left(p^2+m_3^2\right) D_4 + \left(p^2-m_3^2\right)^2 \right].
\eq
For the Baikov representation we do not worry so much about the integration domain, nor about constant prefactors.
For this reason we used a $\sim$-sign instead of an equal sign.
The Baikov representation is particularly suited to obtain the maximal cut of a Feynman integral up to constant prefactors.
The maximal cut of a Feynman integral is obtained by replacing each propagator $1/D_j$ by a Dirac $\delta$-distribution $(2\pi i) \delta(D_j)$.
For the integral $S_{111}(0,x,y_1,y_2)$ one finds
\bq
\label{maxcut}
\lefteqn{
 \mathrm{MaxCut} \; S_{111}\left( 0, x, y_1, y_2 \right)
 \sim } & & \\
 & &
 \frac{m_3^2}{\pi^2}
 \int 
 \frac{dD_4}{\sqrt{ \left[ D_4 + \left(m_1+m_2\right)^2\right] \left[ D_4+\left(m_1-m_2\right)^2\right] \left[D_4^2 + 2 \left(p^2+m_3^2\right) D_4 + \left(p^2-m_3^2\right)^2 \right]}}.
 \nonumber
\eq

\subsection{Definitions}

The sunrise integral as defined in eq.~(\ref{def_momentum_space}) has an obvious symmetry under the symmetric group $S_3$,
which acts by permuting the tuples $(m_1,\nu_1)$, $(m_2,\nu_2)$ and $(m_3,\nu_3)$.
In cases where we want to emphasize this symmetry, we will use the original variables $(p^2,m_1^2,m_2^2,m_3^2)$.
On the other hand, the variables $(x,y_1,y_2)$ are more closely related to coordinates on the moduli space ${\mathcal M}_{1,3}$.

In order to keep the expressions compact, it is convenient to introduce a few abbreviations:
We set
\bq
 t \; = \;p^2.
\eq
We denote the masses related to the pseudo-thresholds by
\bq
\label{def_pseudo_thresholds}
 \mu_1
 =
 -m_1+m_2+m_3,
 \;\;\;
 \mu_2
 =
 m_1-m_2+m_3,
 \;\;\;
 \mu_3
 =
 m_1+m_2-m_3,
\eq
and the mass related to the threshold by
\bq
\label{def_thresholds}
 \mu_4 = m_1+m_2+m_3.
\eq
We introduce the monomial symmetric polynomials $M_{\lambda_1 \lambda_2 \lambda_3}$ in the variables $m_1^2$, $m_2^2$ and $m_3^2$.
These are defined by
\bq
 M_{\lambda_1 \lambda_2 \lambda_3} & = &
 \sum\limits_{\sigma} \left( m_1^2 \right)^{\sigma\left(\lambda_1\right)} \left( m_2^2 \right)^{\sigma\left(\lambda_2\right)} \left( m_3^2 \right)^{\sigma\left(\lambda_3\right)},
\eq
where the sum is over all distinct permutations $\sigma$ of $\left(\lambda_1,\lambda_2,\lambda_3\right)$.
A few examples are
\bq
 M_{100}
 & = &
 m_1^2 + m_2^2 + m_3^2,
 \nonumber \\
 M_{111}
 & = &
 m_1^2 m_2^2 m_3^2,
 \nonumber \\
 M_{210}
 & = &
 m_1^4 m_2^2 + m_2^4 m_3^2 + m_3^4 m_1^2 + m_2^4 m_1^2 + m_3^4 m_2^2 + m_1^4 m_3^2.
\eq
In addition, we introduce the abbreviation
\bq
\label{def_delta}
\Delta & = & \mu_1 \mu_2 \mu_3 \mu_4.
\eq

\subsection{The moduli space}

Let us now consider a smooth algebraic curve $C$ in ${\mathbb C}{\mathbb P}^2$ of genus $g$ with $n$ marked points.
Two such curves $(C;z_1,...,z_n)$ and $(C';z_1',...,z_n')$ are isomorphic if there is an isomorphism
\bq
 \phi & : & C \rightarrow C'
 \;\;\;\;\;\;
 \mbox{such that} \;\;
 \phi\left(z_i\right) = z_i'.
\eq
The moduli space
\bq
 {\mathcal M}_{g,n}
\eq
is the space of isomorphism classes of smooth algebraic curves of genus $g$ with $n$ marked points.
For $g \ge 1$ the isomorphism classes do not only depend on the positions of the marked points,
but also on the ``shape'' of the curve.
For $g=0$ there is only one ``shape'', the Riemann sphere.
The dimension of ${\mathcal M}_{g,n}$ is
\bq
 \dim\left({\mathcal M}_{g,n}\right) & = & 3g + n -3.
\eq
Let us now specialise to $g=1$.
A smooth algebraic curve of genus one with one marked point is an elliptic curve 
and ${\mathcal M}_{1,1}$ parametrises the isomorphism classes of elliptic curves.
We may represent an elliptic curve by
\bq 
 {\mathbb C} / \Lambda,
\eq
with the lattice $\Lambda$ given by ${\mathbb Z} + \tau {\mathbb Z}$.
The marked point is the origin in ${\mathbb C} / \Lambda$.
The moduli space ${\mathcal M}_{1,1}$ has dimension one and we may take the modular parameter $\tau$ as a local coordinate.

Let us now turn to the moduli space ${\mathcal M}_{1,3}$.
As before, one marked point is given by the origin. 
We denote by $z_1, z_2 \in {\mathbb C} / \Lambda$ the other two marked points.
We have $\dim({\mathcal M}_{1,3})=3$ and our standard local coordinates for ${\mathcal M}_{1,3}$ are
\bq
 \left( \tau, z_1, z_2 \right).
\eq
The moduli space ${\mathcal M}_{g,n}$ is not compact. We denote by $\overline{\mathcal M}_{g,n}$
the Deligne-Mumford-Knudsen compactification \cite{Deligne:1969,Knudsen:1976,Knudsen:1983,Knudsen:1983a}.
The space $\overline{\mathcal M}_{g,n}$ is the moduli space of stable nodal curves of arithmetic genus $g$ with
$n$ marked points.
${\mathcal M}_{g,n}$ is an open subset in $\overline{\mathcal M}_{g,n}$.

\section{Elliptic curves}
\label{sec:elliptic_curve}

The sunrise integral with unequal masses is associated to an isogeny class of elliptic curves.
We may think of an elliptic curve as being defined as ${\mathbb C}/\Lambda$, where $\Lambda$ is a lattice.
Two elliptic curves $E = {\mathbb C}/\Lambda$ and $E' = {\mathbb C}/\Lambda'$ are isomorphic, if there is a complex number $c$
such that
\bq
 c \Lambda & = & \Lambda'.
\eq
Two elliptic curves are isogenic, if there is a complex number $c$ such that
\bq
\label{def_isogeny}
 c \Lambda & \subset & \Lambda',
\eq
i.e. $c \Lambda$ is a sub-lattice of $\Lambda'$.

For the sunrise graph there are two obvious ways to associate an elliptic curve to this Feynman integral.
One may either define the elliptic curve from the second graph polynomial
or from the maximal cut.
These two curves are not isomorphic, but isogenic.
Let us denote the lattice for the curve obtained from the Feynman graph polynomial by $\Lambda_F$
and the lattice for the curve obtained from the maximal cut by $\Lambda_C$.
Then $\Lambda_F$ is a sub-lattice of $\Lambda_C$ of index $2$
and -- in view of definition~(\ref{def_isogeny}) -- the lattice $2\Lambda_C$ is a sub-lattice of $\Lambda_F$.
The lattice $\Lambda_F$ is rectangular for $t<0$ and was used in \cite{Adams:2014vja,Adams:2015gva},
while the definition of the elliptic curve from the maximal cuts 
generalises to more complicated integrals \cite{Adams:2018bsn,Adams:2018kez}. 

In appendix \ref{sect:details_elliptic_curves} we give a detailed account on the relation between the two elliptic curves.
In particular, we define the modular parameter $\tau_F$ for the elliptic curve obtained from the Feynman graph polynomial
and the modular parameter $\tau_c$ for the elliptic curve obtained from the maximal cut.
The elliptic curve obtained from the Feynman graph polynomial comes naturally with three marked points: These are the intersections
of the elliptic curve with the Feynman parameter integration region.
On ${\mathbb C}/\Lambda_F$ we denote these points by $0,z_{1,F},z_{2,F}$.
One marked point can be chosen as the origin in ${\mathbb C}/\Lambda_F$.
The corresponding points on ${\mathbb C}/\Lambda_C$ are denoted by $0,z_{1,C},z_{2,C}$.
Appendix \ref{sect:details_elliptic_curves} also provides explicit formulae how the points $z_{i,F}$ and $z_{i,C}$ are related.


\section{Variables}
\label{sect:variables}

The dependence on the kinematics is described by the three variables $(x,y_1,y_2)$.
We will change to new variables $(\tau,z_1,z_2)$, which 
are coordinates on the moduli space ${\mathcal M}_{1,3}$ of a genus one Riemann surface with three marked points.
The new variables are:
\begin{alignat}{2}
\label{trafo_to_moduli_space}
 \tau
 & \; = \;  
 \tau_C
 & \; = \; &
 \frac{\tau_F'}{2}
 \nonumber \\
 z_1
 & \; = \;  
 z_{1,C}
 & \; = \; &
 z_{1,F},
 \nonumber \\
 z_2
 & \; = \; 
 z_{2,C}
 & \; = \; &
 z_{2,F}.
\end{alignat}
Eq.~(\ref{trafo_to_moduli_space}) gives the transformation from the variables $(x,y_1,y_2)$ to the variables $(\tau,z_1,z_2)$.
We are also interested in the inverse transformation.
To find this inverse transformation, we first consider intermediary coordinates $(\lambda,\kappa_1,\kappa_2)$ defined
by
\begin{alignat}{2}
 \lambda 
 & \; = \;  
 k_C^2
 & \; = \; &
 \frac{16 m_1 m_2 m_3 \sqrt{t}}{\left(\mu_1+\sqrt{t}\right)\left(\mu_2+\sqrt{t}\right)\left(\mu_3+\sqrt{t}\right)\left(\mu_4-\sqrt{t}\right)},
 \nonumber \\
 \kappa_1 
 & \; = \;  
 u_{1,C,1/2}^2
 & \; = \; &
 \frac{\left(\mu_2+\sqrt{t}\right)\left(\mu_3+\sqrt{t}\right)}{4 m_2 m_3},
 \nonumber \\
 \kappa_2 
 & \; = \; 
 u_{2,C,1/2}^2
 & \; = \; &
 \frac{\left(\mu_1+\sqrt{t}\right)\left(\mu_3+\sqrt{t}\right)}{4 m_1 m_3}.
\end{alignat}
We may express $(x,y_1,y_2)$ in terms of $(\lambda,\kappa_1,\kappa_2)$:
\bq
\label{lambda_to_x}
 x
 & = &
 \frac{\left(1-\kappa_1\right)\left(1-\kappa_2\right) \kappa_1 \kappa_2 \lambda^2}{\left(1-\lambda\kappa_1\right) \left(1-\lambda\kappa_2\right)},
 \nonumber \\
 y_1
 & = &
 \frac{\kappa_1 \left(1-\kappa_1\right)}{\left(1-\lambda\kappa_1\right) \left(\kappa_1-\kappa_2\right)^2\left(1-\kappa_1-\kappa_2+\lambda\kappa_1\kappa_2\right)^2} R,
 \nonumber \\
 y_2
 & = &
 \frac{\kappa_2 \left(1-\kappa_2\right)}{\left(1-\lambda\kappa_2\right) \left(\kappa_1-\kappa_2\right)^2\left(1-\kappa_1-\kappa_2+\lambda\kappa_1\kappa_2\right)^2} R,
\eq
with
\bq
 R
 & = &
 \left( 1+\kappa_1^{3}\kappa_2^{3}\lambda^{3} \right)  \left( \kappa_1+\kappa_2 \right) 
 - \left( 1+\kappa_1^{2}\kappa_2^{2}\lambda^{2} \right)  \left( \kappa_1^{2}+\kappa_2^{2} \right)  \left( 1+\lambda \right) 
- 8 \left( 1+\kappa_1^{2}\kappa_2^{2}\lambda^{2} \right) \kappa_1\,\kappa_2
 \nonumber \\
 & &
+ \left( 1+\lambda\,\kappa_1\,\kappa_2 \right)  \left( \kappa_1+\kappa_2 \right) ^{3}\lambda
+ 3 \left( 1+\lambda\,\kappa_1\,\kappa_2 \right) \kappa_1\,\kappa_2\, \left( \kappa_1+\kappa_2 \right) \lambda
 \nonumber \\
 & &
+ 8 \left( 1+\lambda\,\kappa_1\,\kappa_2 \right) \kappa_1\,\kappa_2\, \left( \kappa_1+\kappa_2 \right) 
+4\,\kappa_1^{2}\kappa_2^{2} \lambda \left( 1-\lambda \right) 
-8\,\kappa_1\,\kappa_2\, \left( \kappa_1+\kappa_2 \right) ^{2}\lambda-8\,\kappa_1^{2}\kappa_2^{2}
 \nonumber \\
 & &
 - 2 \left(1-2\kappa_1+\lambda\kappa_1^2\right) \left(1-2\kappa_2+\lambda\kappa_2^2\right)
  \sqrt{ \kappa_1 \kappa_2 \left(1-\kappa_1\right)\left(1-\kappa_2\right) \left(1-\lambda\kappa_1\right) \left(1-\lambda\kappa_2\right)}.
 \nonumber \\
\eq
On the other hand we have
\bq
\label{tau_to_lambda}
 \lambda
 & = &
 \frac{\theta_2^4\left(0,q\right)}{\theta_3^4\left(0,q\right)},
 \nonumber \\
 \kappa_1
 & = &
 \frac{\theta_3^2\left(0,q\right)}{\theta_2^2\left(0,q\right)} 
 \frac{\theta_1^2\left(\frac{\pi z_{1}}{2}, q\right)}{\theta_4^2\left(\frac{\pi z_{1}}{2}, q\right)},
 \nonumber \\
 \kappa_2
 & = &
 \frac{\theta_3^2\left(0,q\right)}{\theta_2^2\left(0,q\right)} 
 \frac{\theta_1^2\left(\frac{\pi z_{2}}{2}, q\right)}{\theta_4^2\left(\frac{\pi z_{2}}{2}, q\right)}.
\eq
Eq.~(\ref{lambda_to_x}) and eq.~(\ref{tau_to_lambda}) 
together with $q=\exp(i\pi \tau)$ 
allow us to express $(x,y_1,y_2)$ in terms of $(\tau,z_1,z_2)$.

We set
\begin{alignat}{3}
 q & = \; e^{\pi i \tau},
 &
 \;\;\;\;\;\;
 w_1 & = \; e^{\pi i z_{1}},
 & 
 \;\;\;\;\;\;
 w_2 & = \; e^{\pi i z_{2}},
 \nonumber \\
 \bar{q} & = \; e^{2 \pi i \tau},
 &
 \;\;\;\;\;\;
 \bar{w}_1 & = \; e^{2 \pi i z_{1}},
 &
 \;\;\;\;\;\;
 \bar{w}_2 & = \; e^{2 \pi i z_{2}}.
\end{alignat}
On ${\mathbb C}/\Lambda_F$ we have up to now three marked points: $0$, $z_1$ and $z_2$.
We introduce one further marked point through
\bq
 z_3 & = & 1 - z_1 - z_2
\eq
and set
\bq
 \bar{w}_3 & = & e^{2 \pi i z_{3}} \; = \; \bar{w}_1^{-1} \bar{w}_2^{-1}.
\eq
We further set
\bq
 \psi_1 & = & \psi_{1,C} \; = \; \psi_{1,F} \; = \; \psi_{1,F}',
 \nonumber \\
 W_t & = & 
 \psi_{1,C} \frac{d}{dt} \psi_{2,C} - \psi_{2,C} \frac{d}{dt} \psi_{1,C}
 \; = \;
 \frac{2 \pi i \mu^4 \left(-3t^2 +2 M_{100}t - \Delta \right)}{t \left(t-\mu_1^2\right)\left(t-\mu_2^2\right)\left(t-\mu_3^2\right)\left(t-\mu_4^2\right)}.
\eq


\section{Special functions}
\label{sect:special_functions}

From the Kronecker function $F(x,y,\tau)$ (with $q=\exp(\pi i \tau)$)
\bq
 F\left(x,y,\tau\right)
 & = &
 \pi
 \theta_1'\left(0,q\right) \frac{\theta_1\left( \pi\left(x+y\right), q \right)}{\theta_1\left( \pi x, q \right)\theta_1\left( \pi y, q \right)}
\eq
one defines $g^{(n)}(z,\tau)$ through
\bq
\label{def_g_n}
 F\left(z,\alpha,\tau\right)
 & = &
 \frac{1}{\alpha} \sum\limits_{n=0}^\infty g^{(n)}\left(z,\tau\right) \alpha^n.
\eq
In previous publications we introduced the notation \cite{Adams:2014vja,Adams:2015gva,Adams:2015ydq,Adams:2016xah}
\bq
 \mathrm{ELi}_{n;m}\left(x;y;q\right) & = & 
 \sum\limits_{j=1}^\infty \sum\limits_{k=1}^\infty \; \frac{x^j}{j^n} \frac{y^k}{k^m} q^{j k}.
\eq
We define the linear combinations
\bq
 \overline{\mathrm{E}}_{n;m}\left(x;y;q\right) 
 & = &
  \mathrm{ELi}_{n;m}\left(x;y;q\right)
  - \left(-1\right)^{n+m} \mathrm{ELi}_{n;m}\left(x^{-1};y^{-1};q\right).
\eq
The functions $\mathrm{ELi}_{n;m}$ are helpful for the $\bar{q}$-expansion of the functions $g^{(n)}(z,\tau)$.
Explicitly one has with $\bar{q}=\exp(2\pi i\tau)$ and $\bar{w}=\exp(2\pi i z)$
\bq
\label{g_n_explicit}
 g^{(0)}\left(z,\tau\right)
 & = & 1,
 \nonumber \\
 g^{(1)}\left(z,\tau\right)
 & = &
 - 2 \pi i \left[
                  \frac{1+\bar{w}}{2 \left(1-\bar{w}\right)}
                  + \overline{\mathrm{E}}_{0,0}\left(\bar{w};1;\bar{q}\right)
 \right],
 \nonumber \\
 g^{(n)}\left(z,\tau\right)
 & = &
 - \frac{\left(2\pi i\right)^n}{\left(n-1\right)!} 
 \left[
 - \frac{B_n}{n}
       + \overline{\mathrm{E}}_{0,1-n}\left(\bar{w};1;\bar{q}\right)
 \right],
 \;\;\;\;\;\;\;\;\;\;\;\;\;\;\;\;\;\;\;\;\;\;\;\;\;\;\;
 n > 1,
\eq
where $B_n$ denote the Bernoulli numbers, defined by
\bq
 \frac{x}{e^x-1}
 & = &
 \sum\limits_{n=0}^\infty \frac{B_n}{n!} x^n.
\eq
The functions $g^{(n)}(z,\tau)$ will play prominent roles in our calculation.
Let us therefore discuss their properties \cite{Brown:2011,Broedel:2018qkq}.
\\
\\
{\bf Poles}: When viewed as a function of $z$, the function $g^{(n)}(z,\tau)$ has only simple poles.
More concretely, the function $g^{(1)}(z,\tau)$ has a simple pole with unit residue at every point of the lattice.
For $n>1$ the function $g^{(n)}(z,\tau)$ has a simple pole only at those lattice points that do not lie on the real axis.
\\
\\
{\bf Periodicity}:
The (quasi-) periodicity properties are
\bq
 g^{(n)}\left(z+1,\tau\right) & = &  g^{(n)}\left(z,\tau\right),
 \nonumber \\
 g^{(n)}\left(z+\tau,\tau\right) & = &  
 \sum\limits_{j=0}^n \frac{\left(-2\pi i\right)^j}{j!} g^{(n-j)}\left(z,\tau\right).
\eq
We see that $g^{(n)}(z,\tau)$ is invariant under translations by $1$, but not by $\tau$.
Thus we may view $g^{(n)}(z,\tau)$ either as a multi-valued function 
on the moduli space of a genus one Riemann surface with
marked points or alternatively as a single-valued function 
on the covering space of the moduli space of a genus one Riemann surface with
marked points.

In the following we will take $g^{(n)}(z,\tau)$ as a multi-valued function 
on the moduli space of a genus one Riemann surface with
marked points and we will be working with the branch defined by eq.~(\ref{g_n_explicit}).
\\
\\
{\bf Modularity}: Let us now consider
\bq
 g^{(n)}\left( \frac{z}{c\tau +d}, \frac{a\tau+b}{c\tau+d} \right),
 & &
 \left( \begin{array}{cc} 
       a & b \\
       c & d \\
 \end{array} \right)
 \; \in \; \mathrm{SL}_2\left({\mathbb Z}\right). 
\eq
The functions $g^{(n)}(z,\tau)$ may be expressed in terms of Eisenstein series, and we may derive the behaviour of $g^{(n)}(z,\tau)$
under modular transformations from those.
The Eisenstein series are defined by
\bq
 E_k\left(z,\tau\right)
 & = &
 \sideset{}{_e}\sum\limits_{(n_1,n_2) \in {\mathbb Z}^2} \frac{1}{\left(z+n_1 + n_2\tau \right)^k}.
\eq
The series is absolutely convergent for $k \ge 3$.
For $k=1$ and $k=2$ the Eisenstein summation depends on the choice of generators. The Eisenstein summation prescription is defined by
\bq
 \sideset{}{_e}\sum\limits_{(n_1,n_2) \in {\mathbb Z}^2} f\left(z+n_1 + n_2\tau \right)
 & = &
 \lim\limits_{N_2\rightarrow \infty} \sum\limits_{n_2=-N_2}^{N_2}
 \left(
 \lim\limits_{N_1\rightarrow \infty} \sum\limits_{n_1=-N_1}^{N_1}
 f\left(z + n_1 + n_2 \tau \right)
 \right).
\eq
One further sets
\bq
 e_k\left(\tau\right)
 & = &
 \sideset{}{_e}\sum\limits_{(n_1,n_2) \in {\mathbb Z}^2\backslash (0,0)} \frac{1}{\left(n_1 + n_2\tau \right)^k}.
\eq
An alternative definition of the Kronecker function is given in terms of Eisenstein series:
\bq
 F\left(x,y,\tau\right)
 & = &
 \frac{1}{y} \exp\left(- \sum\limits_{k=1}^\infty \frac{\left(-y\right)^k}{k} \left( E_k\left(x,\tau\right) - e_k\left(\tau\right) \right) \right).
\eq
This allows us to express the functions $g^{(n)}(z,\tau)$ in terms of Eisenstein series.
The explicit expressions for the first five functions $g^{(n)}(z,\tau)$ in terms of Eisenstein series read
\bq
 g^{(0)}\left(z,\tau\right)
 & = & 
 1,
 \nonumber \\
 g^{(1)}\left(z,\tau\right)
 & = & 
 E_1\left(z,\tau\right),
 \nonumber \\
 g^{(2)}\left(z,\tau\right)
 & = & 
 - \frac{1}{2} \left[ E_2\left(z,\tau\right) - e_2\left(\tau\right) - E_1\left(z,\tau\right)^2 \right],
 \nonumber \\
 g^{(3)}\left(z,\tau\right)
 & = & 
 \frac{1}{6} \left\{ 2 E_3\left(z,\tau\right) - 3 \left[ E_2\left(z,\tau\right) - e_2\left(\tau\right) \right] E_1\left(z,\tau\right) + E_1\left(z,\tau\right)^3 \right\},
 \nonumber \\
 g^{(4)}\left(z,\tau\right)
 & = & 
 - \frac{1}{24} \left\{ 6 \left[ E_4\left(z,\tau\right) - e_4\left(\tau\right) \right] - 8 E_3\left(z,\tau\right) E_1\left(z,\tau\right)
                      - 3 \left[ E_2\left(z,\tau\right) - e_2\left(\tau\right) \right]^2
 \right. \nonumber \\
 & & \left.
                      + 6 \left[ E_2\left(z,\tau\right) - e_2\left(\tau\right) \right] E_1\left(z,\tau\right)^2
                      - E_1\left(z,\tau\right)^4 \right\}.
\eq
For $k \ge 3$ the Eisenstein series transform nicely under modular transformations.
With 
\bq
 \tau' \; = \; \frac{a\tau+b}{c\tau+d},
 \;\;\;\;\;\;
 z' \; = \; \frac{z}{c\tau+d},
 \;\;\;\;\;\;
 \gamma \; = \;
 \left( \begin{array}{cc} 
       a & b \\
       c & d \\
 \end{array} \right)
 \; \in \; \mathrm{SL}_2\left({\mathbb Z}\right)
\eq
one has
\bq
 E_k\left(z',\tau'\right) \; = \; \left(c\tau+d\right)^k E_k\left(z,\tau\right),
 & &
 e_k\left(\tau'\right) \; = \; \left(c\tau+d\right)^k e_k\left(\tau\right).
\eq
This leaves the cases $k=1$ and $k=2$.
At modular weight $k=2$ only the combination
\bq
 E_2\left(z,\tau\right) - e_2\left(\tau\right)
 & = &
 \wp\left(z\right)
\eq
enters.
This combination (but not the individual terms) transforms nicely:
\bq
\label{modular_trafo_weight_2}
 E_2\left(z',\tau'\right) - e_2\left(\tau'\right)
 & = &
 \left(c\tau +d\right)^2 \left[ E_2\left(z,\tau\right) - e_2\left(\tau\right) \right].
\eq
The nice transformation properties are spoiled by $E_1\left(z,\tau\right)$.
One has
\bq
\label{modular_trafo_weight_1}
 E_1\left(z',\tau'\right)
 & = &
 \left(c\tau +d\right) \left[ E_1\left(z,\tau\right) + C_1 z + C_0 \right],
\eq
where the constants $C_1$ and $C_0$ depend on $\gamma$.
\\
\\
{\bf Differential forms}: Let us set $g^{(-1)}(z,\tau)=0$.
We define differential forms by
\bq
\label{def_omega}
 \omega_k\left(z,\tau\right)
 & = &
 \left(2\pi\right)^{2-k}
 \left[
  g^{(k-1)}\left(z,\tau\right) dz + \left(k-1\right) g^{(k)}\left(z,\tau\right) \frac{d\tau}{2\pi i}
 \right].
\eq
The differential forms $\omega_k(z,\tau)$ appear in the total differential of elliptic polylogarithms \cite{Broedel:2018iwv}.
A few examples are (recall $z_3=1-z_1-z_2$):
\bq
 \omega_0\left(z_i,\tau\right)
 & = &
 2 \pi i d\tau,
 \nonumber \\
 \omega_1\left(z_1,\tau\right)
 & = &
 2 \pi dz_1,
 \nonumber \\
 \omega_1\left(z_3,\tau\right)
 & = &
 - 2 \pi dz_1 - 2 \pi dz_2,
 \nonumber \\
 \omega_2\left(z_1,\tau\right)
 & = &
 g^{(1)}\left(z_1,\tau\right) dz_1 + g^{(2)}\left(z_1,\tau\right) \frac{d\tau}{2\pi i},
 \nonumber \\
 \omega_2\left(z_1,2\tau\right)
 & = &
 g^{(1)}\left(z_1,2\tau\right) dz_1 + 2 g^{(2)}\left(z_1,2\tau\right) \frac{d\tau}{2\pi i}.
\eq
Since $\omega_0(z,\tau)$ does not depend on $z$ and $\omega_1(z,\tau)$ does not depend on $\tau$
we write
\bq
 \omega_0\left(\tau\right)
 \; = \;
 \omega_0\left(z,\tau\right),
 & &
 \omega_1\left(z\right)
 \; = \;
 \omega_1\left(z,\tau\right).
\eq
In addition we need two differential forms $\eta_2(\tau)$ and $\eta_4(\tau)$, which depend on $\tau$, but not
on the $z_i$'s.
The first one is related to a modular form of $\Gamma_0(2)$, the second one to a modular form of $\mathrm{SL}_2({\mathbb Z})$.
Let $b_2(\tau)$ be the generator of ${\mathcal M}_2(\Gamma_0(2))$ given in terms of the Eisenstein series defined above as
\bq
 b_2\left(\tau\right)
 & = &
 e_2\left(\tau\right) - 2  e_2\left(2\tau\right).
\eq
The first few terms of the $\bar{q}$-expansion read
\bq
 b_2\left(\tau\right)
 & = &
 2 \left(2 \pi i \right)^2 \left[ 
 \frac{1}{24} + \bar{q} + \bar{q}^2 + 4 \bar{q}^3 + \bar{q}^4 + 6 \bar{q}^5 + ... 
\right].
\eq
We set
\bq
 \eta_2\left(\tau\right)
 \; = \;
 b_2\left(\tau\right) \frac{d\tau}{2\pi i},
 & &
 \eta_4\left(\tau\right)
 \; = \;
 \frac{1}{\left(2\pi\right)^2} e_4\left(\tau\right) \frac{d\tau}{2\pi i}.
\eq


\section{Master integrals}
\label{sect:masters}

In this section we present a basis $\vec{J}$ of master integrals for the unequal mass sunrise family, such that 
the associated differential equation is in $\eps$-form.
The associated differential equation is given in the next section.

The starting point is a pre-canonical basis of master integrals.
A possible pre-canonical basis is given by 
\bq
 \vec{I} & = &
 \left(
 S_{110},
 \;
 S_{101},
 \;
 S_{011},
 \;
 S_{111},
 \;
 S_{211},
 \;
 S_{121},
 \;
 S_{112}
 \right)^T.
\eq
The associated differential equation is of the form
\bq
 d \vec{I} & = & \tilde{A}\left(\eps,x,y_1,y_2\right) \vec{I},
 \nonumber \\
 \tilde{A}\left(\eps,x,y_1,y_2\right) & = &
 \tilde{A}_x\left(\eps,x,y_1,y_2\right) dx
 +
 \tilde{A}_{y_1}\left(\eps,x,y_1,y_2\right) dy_1
 +
 \tilde{A}_{y_2}\left(\eps,x,y_1,y_2\right) dy_2.
\eq
We then change the basis of master integrals
\bq
\label{fibre_trafo}
 \vec{J} & = & U \vec{I}
\eq
and the coordinates on the base manifold
\bq
\label{base_trafo}
 \left(x,y_1,y_2\right) & \rightarrow & \left(\tau,z_1,z_2\right).
\eq
Under the transformation~(\ref{fibre_trafo}) the differential equation transforms into
\bq
 d \vec{J} & = & \hat{A} \vec{J},
\eq
where $\hat{A}$ is related to $\tilde{A}$ by
\bq
 \hat{A}
 & = &
 U \tilde{A} U^{-1}
 -
 U d U^{-1}.
\eq
A change of coordinates on the base manifold leads to the transformation
\bq
 \hat{A}
 & = &
 \hat{A}_x dx
 +
 \hat{A}_{y_1} dy_1
 +
 \hat{A}_{y_2} dy_2
 \; = \;
 \hat{A}_\tau d\tau
 +
 \hat{A}_{z_1} dz_1
 +
 \hat{A}_{z_2} dz_2,
\eq
where for example $\hat{A}_\tau$ is related to $\hat{A}_x$, $\hat{A}_{y_1}$, $\hat{A}_{y_2}$ by
\bq
 \hat{A}_\tau
 & = &
 \hat{A}_x \frac{\partial x}{\partial \tau}
 + \hat{A}_{y_1} \frac{\partial y_1}{\partial \tau}
 + \hat{A}_{y_2} \frac{\partial y_2}{\partial \tau}.
\eq
The essential step is to find a matrix $U$ such that
\bq
 \hat{A} & = & \eps A,
\eq
where $A$ does not depend on $\eps$.
We construct $U$ in two steps:
\bq
 U & = & U_2 U_1,
\eq
Let us set
\bq
 \vec{J}_1 \; = \; U_1 \vec{I},
 & &
 d \vec{J}_1 \; = \; \hat{A}_1 \vec{J}_1.
\eq
The entries of $U_1$ are constructed such that $\hat{A}_1$ is linear in $\eps$ and the $\eps^0$-part is strictly lower triangular, i.e.
\bq
 \hat{A}_1 & = & \hat{A}^{(0)}_1 + \eps \hat{A}^{(1)}_1,
\eq
where $\hat{A}^{(0)}_1$ and $\hat{A}^{(1)}_1$ are independent of $\eps$ and
$\hat{A}^{(0)}_1$ is strictly lower triangular.
This can be done with a transformation, where the entries of $U_1$ are rational functions of $\eps$, $x$, $y_1$, $y_2$, $\psi_1$ and $\partial_x \psi_1$ (i.e. compared
to the full transformation $U$ the entries of $U_1$ do not involve incomplete elliptic integrals. They only involve complete elliptic integrals
related to $\psi_1$ and $\partial_x \psi_1$).
The entries of $U_1$ are determined from an ansatz, following the ideas of \cite{Meyer:2016slj,Meyer:2017joq,Adams:2018bsn,Adams:2018kez}.
Explicitly, $U_1$ is given by setting $F_{54}=F_{64}=F_{74}=0$ in the formula~(\ref{def_J}) below.

In a second step $U_2$ is constructed. $U_2$ eliminates the non-zero entries of $\hat{A}^{(0)}_1$. As $\hat{A}^{(0)}_1$
is strictly lower triangular, this can be done systematically by integration and will lead to incomplete elliptic integrals.
In a final clean-up (and after the change of coordinates on the base manifold) we trade the incomplete elliptic integrals
for $d\ln(y_1)/d\tau$ and $d\ln(y_2)/d\tau$.

Let us now present the basis $\vec{J}$:
\bq
\label{def_J}
 J_1
 & = &
 \eps^2 S_{101},
 \nonumber \\
 J_2
 & = &
 \eps^2 S_{011},
 \nonumber \\
 J_3
 & = &
 \eps^2 S_{110},
 \nonumber \\
 J_4 
 & = &
 \eps^2 \frac{\pi}{\psi_1} S_{111},
 \nonumber \\
 J_5 
 & = &
 \eps \left[ 
             \frac{\left(m_1^2+m_2^2-2m_3^2\right)}{\mu^2} S_{111}
             + \frac{\left(t-m_1^2-3m_2^2+3m_3^2\right) m_1^2}{\mu^4} S_{211}
 \right. \nonumber \\
 & & \left.
             + \frac{\left(t-3m_1^2-m_2^2+3m_3^2\right) m_2^2}{\mu^4} S_{121}
             - \frac{2 \left(t-m_3^2\right) m_3^2}{\mu^4} S_{112}
      \right]
 + \frac{2 \eps^2}{\left( 3 t^2-2 M_{100} t + \Delta \right) \mu^2}
 \nonumber \\
 & &
 \times
   \left[
          7 \left(m_1^2+m_2^2-2m_3^2\right) t^2 
          - 2 \left(3m_1^4+3m_2^4 -6m_3^4 + m_1^2 m_3^2 + m_2^2 m_3^2 - 2 m_1^2 m_2^2 \right) t
 \right. \nonumber \\
 & & \left.
          + \left(m_1^2+m_2^2 - 2 m_3^2 \right) \Delta
   \right]
   S_{111}
 + F_{54} J_4,
 \nonumber \\
 J_6 
 & = &
 \eps \left[ 
             \frac{\left(m_1^2-m_2^2\right)}{\mu^2} S_{111}
             + \frac{\left(t-m_1^2+m_2^2-m_3^2\right) m_1^2}{\mu^4} S_{211}
             - \frac{\left(t+m_1^2-m_2^2-m_3^2\right) m_2^2}{\mu^4} S_{121}
 \right. \nonumber \\
 & & \left.
             - \frac{2 \left(m_1^2-m_2^2\right) m_3^2}{\mu^4} S_{112}
      \right]
 + \frac{2 \eps^2 \left(m_1^2-m_2^2\right) }{\left( 3 t^2-2 M_{100} t + \Delta \right) \mu^2}
   \left[
          7 t^2 
          - 2 \left(3m_1^2+3m_2^2 -m_3^2 \right) t 
          + \Delta
   \right]
   S_{111}
 \nonumber \\
 & &
 + F_{64} J_4,
 \nonumber \\
 J_7 
 & = &
 \frac{1}{\eps} \frac{\psi_1^2}{2\pi i W_t} \frac{d}{dt} J_4
 + \frac{\eps^2}{8}  \frac{1}{\left(3t^2-2M_{100}t+\Delta \right)^2 \mu^4} 
 \left[ 
        9 t^6
        - 22 M_{100} t^5
        + \left( 50 M_{110} - M_{200} \right) t^4
 \right. \nonumber \\
 & & \left.
        + \left( 44 M_{300} - 76 M_{210} 
        + 216 M_{111} \right) t^3
        + \left( -41 M_{400} + 84 M_{310} - 86 M_{220} - 52 M_{211} \right) t^2
 \right. \nonumber \\
 & & \left.
        + 2 \Delta \left( - 5 M_{300} + 5 M_{210} 
        - 2 M_{111} \right) t
        - \Delta^3
 \right] \frac{\psi_1}{\pi} S_{111}
 - \frac{1}{8} F_{64} J_6
 - \frac{1}{24} F_{54} J_5
 + F_{74} J_4.
\eq
The three functions $F_{54}$, $F_{64}$, $F_{74}$, appearing in the definition of $J_5$, $J_6$ and $J_7$
are given by
\bq
\lefteqn{
 F_{54}
 = } & &
 \nonumber \\
 & &
 \frac{6 i \mu^2}{\left(3t^2-2M_{100}t+\Delta\right) \psi_1}
 \left[
       \left(m_1^2-m_2^2+m_3^2 - t\right) \frac{1}{y_1} \frac{dy_1}{d\tau}
       +
       \left(-m_1^2+m_2^2+m_3^2 - t\right) \frac{1}{y_2} \frac{dy_2}{d\tau}
 \right],
 \nonumber \\
 & & \nonumber \\
\lefteqn{
 F_{64}
 = } & &
 \nonumber \\
 & &
 \frac{2 i \mu^2}{\left(3t^2-2M_{100}t+\Delta\right) \psi_1}
 \left[
       \left(3m_1^2+m_2^2-m_3^2 - 3t\right) \frac{1}{y_1} \frac{dy_1}{d\tau}
       -
       \left(m_1^2+3m_2^2-m_3^2 - 3t\right) \frac{1}{y_2} \frac{dy_2}{d\tau}
 \right],
 \nonumber \\
 & & \nonumber \\
\lefteqn{
 F_{74}
 = } & &
 \nonumber \\
 & &
 - \frac{\mu^4}{\left(3t^2-2M_{100}t+\Delta\right)^2 \psi_1^2}
 \left[
  \left( 3 m_1^4 + m_2^4 + m_3^4 - 2 m_2^2 m_3^2 - 6 m_1^2 t + 3 t^2 \right) \left(\frac{1}{y_1} \frac{dy_1}{d\tau}\right)^2
 \right. \nonumber \\
 & & \left.
  - \left( 3 m_1^4 + 3 m_2^4 - m_3^4 + 2 m_1^2 m_2^2 - 2 m_1^2 m_3^2 - 2 m_2^2 m_3^2 - 6 \left( m_1^2 +m_2^2 - m_3^2 \right) t + 3 t^2 \right) 
    \left( \frac{1}{y_1} \frac{dy_1}{d\tau} \right) 
 \right. \nonumber \\
 & & \left.
    \times
    \left( \frac{1}{y_2} \frac{dy_2}{d\tau} \right)
  + \left( m_1^4 + 3 m_2^4 + m_3^4 - 2 m_1^2 m_3^2 - 6 m_2^2 t + 3 t^2 \right) \left( \frac{1}{y_2} \frac{dy_2}{d\tau} \right)^2
 \right].
\eq
In the case $m_1=m_2 \neq m_3$ 
we have $J_1=J_2$ and $J_6=0$.
The system reduces to five master integrals.
In the case $m_1=m_2=m_3$ 
we have $J_1=J_2=J_3$ and $J_5=J_6=0$.
The system reduces to three master integrals.


\section{The system of differential equations}
\label{sect:differential_equation}

In the basis $\vec{J}$ the system of differential equations is in $\eps$-form:
\bq
\label{differential_equation_eps_form}
 d \vec{J}
 & = & 
 \eps A \vec{J},
\eq
where the $7 \times 7$-matrix $A$ is independent of $\eps$.
The entries of $A$ are linear combinations of the differential forms $\omega_k$ and $\eta_k$,
defined in section~\ref{sect:special_functions}.
We call $k$ the (generalised) modular weight.
Each entry of the matrix $A$ has a unique modular weight.
In detail, we find that the entries have the following modular weights:
\bq
 \left( \begin{array}{ccccccc}
 2 & - & - & - & - & - & - \\
 - & 2 & - & - & - & - & - \\
 - & - & 2 & - & - & - & - \\
 - & - & - & 2 & 1 & 1 & 0 \\
 2 & 2 & 2 & 3 & 2 & 2 & 1 \\
 2 & 2 & 2 & 3 & 2 & 2 & 1 \\
 3 & 3 & 3 & 4 & 3 & 3 & 2 \\
 \end{array} \right)
\eq
A dash indicates that the corresponding entry is vanishing. Let us now give the non-vanishing entries:
\\
\\
Modular weight $0$:
\bq
 A_{4,7}
 & = &
 \omega_0\left(\tau\right).
\eq
Modular weight $1$:
\bq
 A_{4,5}
 & = &
 \frac{1}{4} \omega_1\left(z_1\right) + \frac{1}{4} \omega_1\left(z_2\right), 
 \nonumber \\
 A_{4,6}
 & = &
 \frac{1}{4} \omega_1\left(z_1\right) - \frac{1}{4} \omega_1\left(z_2\right), 
 \nonumber \\
 A_{5,7}
 & = &
 6 \omega_1\left(z_1\right) + 6 \omega_1\left(z_2\right), 
 \nonumber \\
 A_{6,7}
 & = &
 2 \omega_1\left(z_1\right) - 2 \omega_1\left(z_2\right).
\eq
Modular weight $2$:
\bq
 A_{1,1}
 & = &
 2 \omega_2\left(z_1,\tau\right) - 2 \omega_2\left(z_3,\tau\right) - 4 \omega_2\left(z_1,2\tau\right) + 4 \omega_2\left(z_3,2\tau\right),
 \nonumber \\
 A_{2,2}
 & = &
 2 \omega_2\left(z_2,\tau\right) - 2 \omega_2\left(z_3,\tau\right) - 4 \omega_2\left(z_2,2\tau\right) + 4 \omega_2\left(z_3,2\tau\right),
 \nonumber \\
 A_{3,3}
 & = &
 2 \omega_2\left(z_1,\tau\right) + 2 \omega_2\left(z_2,\tau\right) - 4 \omega_2\left(z_3,\tau\right) - 4 \omega_2\left(z_1,2\tau\right) - 4 \omega_2\left(z_2,2\tau\right) + 8 \omega_2\left(z_3,2\tau\right),
 \nonumber \\
 A_{4,4}
 & = &
 - 4 \omega_2\left(z_3,\tau\right) - 2 \omega_2\left(z_1,2\tau\right) - 2 \omega_2\left(z_2,2\tau\right) + 6 \omega_2\left(z_3,2\tau\right) + 6 \eta_2\left(\tau\right),
 \nonumber \\
 A_{5,1}
 & = &
 2 \omega_2\left(z_1,\tau\right) - 2 \omega_2\left(z_2,\tau\right) - 4 \omega_2\left(z_3,\tau\right) - 4 \omega_2\left(z_1,2\tau\right) + 4 \omega_2\left(z_2,2\tau\right) + 8 \omega_2\left(z_3,2\tau\right),
 \nonumber \\
 A_{5,2}
 & = &
 - 2 \omega_2\left(z_1,\tau\right) + 2 \omega_2\left(z_2,\tau\right) - 4 \omega_2\left(z_3,\tau\right) + 4 \omega_2\left(z_1,2\tau\right) - 4 \omega_2\left(z_2,2\tau\right) + 8 \omega_2\left(z_3,2\tau\right),
 \nonumber \\
 A_{5,3}
 & = &
 2 \omega_2\left(z_1,\tau\right) + 2 \omega_2\left(z_2,\tau\right) + 4 \omega_2\left(z_3,\tau\right) - 4 \omega_2\left(z_1,2\tau\right) - 4 \omega_2\left(z_2,2\tau\right) - 8 \omega_2\left(z_3,2\tau\right),
 \nonumber \\
 A_{5,5}
 & = &
 \omega_2\left(z_1,\tau\right) + \omega_2\left(z_2,\tau\right) - 2 \omega_2\left(z_1,2\tau\right) - 2 \omega_2\left(z_2,2\tau\right) + 6 \omega_2\left(z_3,2\tau\right) + 6 \eta_2\left(\tau\right),
 \nonumber \\
 A_{5,6}
 & = &
 3 \omega_2\left(z_1,\tau\right) - 3 \omega_2\left(z_2,\tau\right),
 \nonumber \\
 A_{6,1}
 & = &
 2 \omega_2\left(z_1,\tau\right) + 2 \omega_2\left(z_2,\tau\right) - 4 \omega_2\left(z_1,2\tau\right) - 4 \omega_2\left(z_2,2\tau\right),
 \nonumber \\
 A_{6,2}
 & = &
 - 2 \omega_2\left(z_1,\tau\right) - 2 \omega_2\left(z_2,\tau\right) + 4 \omega_2\left(z_1,2\tau\right) + 4 \omega_2\left(z_2,2\tau\right),
 \nonumber \\
 A_{6,3}
 & = &
 2 \omega_2\left(z_1,\tau\right) - 2 \omega_2\left(z_2,\tau\right) - 4 \omega_2\left(z_1,2\tau\right) + 4 \omega_2\left(z_2,2\tau\right),
 \nonumber \\
 A_{6,5}
 & = &
 \omega_2\left(z_1,\tau\right) - \omega_2\left(z_2,\tau\right),
 \nonumber \\
 A_{6,6}
 & = &
 3 \omega_2\left(z_1,\tau\right) + 3 \omega_2\left(z_2,\tau\right) - 4 \omega_2\left(z_3,\tau\right) - 2 \omega_2\left(z_1,2\tau\right) - 2 \omega_2\left(z_2,2\tau\right) + 6 \omega_2\left(z_3,2\tau\right) 
 \nonumber \\
 & & 
 + 6 \eta_2\left(\tau\right),
 \nonumber \\
 A_{7,7}
 & = &
 - 4 \omega_2\left(z_3,\tau\right) - 2 \omega_2\left(z_1,2\tau\right) - 2 \omega_2\left(z_2,2\tau\right) + 6 \omega_2\left(z_3,2\tau\right) + 6 \eta_2\left(\tau\right).
\eq
Modular weight $3$:
\bq
 A_{5,4}
 & = &
 - 12 \omega_3\left(z_1,\tau\right) - 12 \omega_3\left(z_2,\tau\right) + 24 \omega_3\left(z_3,\tau\right),
 \nonumber \\
 A_{6,4}
 & = &
 - 12 \omega_3\left(z_1,\tau\right) + 12 \omega_3\left(z_2,\tau\right),
 \nonumber \\
 A_{7,1}
 & = &
 - \omega_3\left(z_1,\tau\right) + \omega_3\left(z_2,\tau\right) - \omega_3\left(z_3,\tau\right) + 4 \omega_3\left(z_1,2\tau\right) - 4 \omega_3\left(z_2,2\tau\right) + 4 \omega_3\left(z_3,2\tau\right),
 \nonumber \\
 A_{7,2}
 & = &
 \omega_3\left(z_1,\tau\right) - \omega_3\left(z_2,\tau\right) - \omega_3\left(z_3,\tau\right) - 4 \omega_3\left(z_1,2\tau\right) + 4 \omega_3\left(z_2,2\tau\right) + 4 \omega_3\left(z_3,2\tau\right),
 \nonumber \\
 A_{7,3}
 & = &
 - \omega_3\left(z_1,\tau\right) - \omega_3\left(z_2,\tau\right) + \omega_3\left(z_3,\tau\right) + 4 \omega_3\left(z_1,2\tau\right) + 4 \omega_3\left(z_2,2\tau\right) - 4 \omega_3\left(z_3,2\tau\right),
 \nonumber \\
 A_{7,5}
 & = &
 - \frac{1}{2} \omega_3\left(z_1,\tau\right) - \frac{1}{2} \omega_3\left(z_2,\tau\right) + \omega_3\left(z_3,\tau\right),
 \nonumber \\
 A_{7,6}
 & = &
 - \frac{3}{2} \omega_3\left(z_1,\tau\right) + \frac{3}{2} \omega_3\left(z_2,\tau\right).
\eq
Modular weight $4$:
\bq
 A_{7,4}
 & = &
 12 \omega_4\left(z_1,\tau\right) + 12 \omega_4\left(z_2,\tau\right) + 12 \omega_4\left(z_3,\tau\right) - 72 \eta_4\left(\tau\right).
\eq
The integration kernels (i.e. the entries of the matrix $A$) are linear combinations of the differential 
one-forms
\bq
 \omega_k\left(z_j, N \tau \right),
 & &
 0 \le k \le 4, 
 \;\;\;
 1 \le j \le 3,
 \;\;\;
 1 \le N \le 2,
\eq
and $\eta_2(\tau)$, $\eta_4(\tau)$.
Let us discuss the differential one-forms $\omega_k(z_j,N\tau)$ as functions of $z_j$.
We recall that $\omega_k$ is defined in eq.~(\ref{def_omega}) with the help of the functions $g^{(k)}(z_j,N\tau)$.
The latter have only single poles, which are all located on the lattice points.
It follows that $\omega_k(z_j,N\tau)$ has only single poles.

The coefficients of the $\eps$-expansion of the master integrals $J_1$-$J_7$ are therefore pure
functions.
We recall that a function is said to be pure
if it is unipotent and its total differential involves only pure functions 
and one-forms with at most logarithmic singularities \cite{Broedel:2018qkq}.
Unipotent means that the function satisfies a differential equation without homogeneous term.
A differential equation in $\eps$-form implies unipotency.
 

\section{Iterated integrals}
\label{sect:iterated_integrals}

Let us review Chen's definition of iterated integrals \cite{Chen}:
Let $M$ be a $n$-dimensional manifold and
\bq
 \gamma & : & \left[a,b\right] \rightarrow M
\eq
a path with start point ${x}_i=\gamma(a)$ and end point ${x}_f=\gamma(b)$.
Suppose further that $\omega_1$, ..., $\omega_k$ are differential $1$-forms on $M$.
Let us write
\bq
 f_j\left(\lambda\right) d\lambda & = & \gamma^\ast \omega_j
\eq
for the pull-backs to the interval $[a,b]$.
For $\lambda \in [a,b]$ the $k$-fold iterated integral 
of $\omega_1$, ..., $\omega_k$ along the path $\gamma$ is defined
by
\bq
 I_{\gamma}\left(\omega_1,...,\omega_k;\lambda\right)
 & = &
 \int\limits_a^{\lambda} d\lambda_1 f_1\left(\lambda_1\right)
 \int\limits_a^{\lambda_1} d\lambda_2 f_2\left(\lambda_2\right)
 ...
 \int\limits_a^{\lambda_{k-1}} d\lambda_k f_k\left(\lambda_k\right).
\eq
We define the $0$-fold iterated integral to be
\bq
 I_{\gamma}\left(;\lambda\right)
 & = &
 1.
\eq
We have
\bq
 \frac{d}{d\lambda}
 I_{\gamma}\left(\omega_1,\omega_2,...,\omega_k;\lambda\right)
 & = &
 f_1\left(\lambda\right) \;
 I_{\gamma}\left(\omega_2,...,\omega_k;\lambda\right).
\eq
Quite often we will be integrating in the variable $\tau$ from $\tau_i=i \infty$ to $\tau_f=\tau$. 
We write 
\bq
 F\left(\omega_1,\omega_2,...,\omega_k\right)
 & = &
 \int\limits_{i \infty}^{\tau}
 \omega_1\left(\tau_1\right)
 \int\limits_{i \infty}^{\tau_1} 
 \omega_2\left(\tau_2\right)
 ...
 \int\limits_{i \infty}^{\tau_{k-1}}
 \omega_k\left(\tau_k\right)
\eq
for the iterated integrals in this case.


\section{Analytical results}
\label{sect:analytical_results}

With the help of the differential equation eq.~(\ref{differential_equation_eps_form})
in $\eps$-form we easily obtain the analytic
solution for the master integrals $J_1$-$J_7$ 
order-by-order in $\eps$ as iterated integrals involving the differential one-forms appearing
in the matrix $A$.
We integrate the differential equation from a chosen boundary point to the desired point in
$\overline{\mathcal M}_{1,3}$.
There is some freedom in choosing the boundary point and the path of integration.
We discuss two possibilities: 
The first option consists in integrating along $\tau$, keeping $z_1$ and $z_2$ constant.
Within the second option we integrate in the $z_1$-$z_2$ subspace, keeping $\tau$ constant.

The first three master integrals $J_1$-$J_3$ are tadpole integrals.
They are given to all order in $\eps$ in the variables $y_1$ and $y_2$ by
\bq
\label{result_tadpoles}
 J_1 & = &
 e^{2 \Eulerconstant \eps} \left( \Gamma\left(1+\eps\right) \right)^2 y_1^{-\eps},
 \nonumber \\
 J_2 & = &
 e^{2 \Eulerconstant \eps} \left( \Gamma\left(1+\eps\right) \right)^2 y_2^{-\eps},
 \nonumber \\
 J_3 & = &
 e^{2 \Eulerconstant \eps} \left( \Gamma\left(1+\eps\right) \right)^2 y_1^{-\eps} y_2^{-\eps}.
\eq

\subsection{Integration along $z_1=\mathrm{const}$ and $z_2=\mathrm{const}$}
\label{subsect:tau_integration}

We may integrate the differential equation along $\tau$, keeping the two other variables $z_1$ and $z_2$
constant:
\bq
 z_1=\mathrm{const}, & & z_2=\mathrm{const}.
\eq
This will give us the closest relation with the equal mass case, where we expressed the equal mass sunrise integral
as iterated integrals of modular forms \cite{Adams:2017ejb,Adams:2018yfj}.
Iterated integrals of modular forms are iterated integrals in the variable $\tau$.
In the equal mass case we have for the variables $z_1$ and $z_2$ for all values of $\tau$
\bq
 z_1^{\mathrm{equal \; mass}} \;\; = \;\;
 z_2^{\mathrm{equal \; mass}} \;\; = \;\;
 \frac{1}{3}.
\eq
Integrating the differential equation along $\tau$ with $z_1=\mathrm{const}$ and $z_2=\mathrm{const}$
requires boundary values for an initial point $\tau_0$.
It is convenient to choose $\tau_0=i\infty$, corresponding to $\bar{q}_0=0$ or $p^2=0$.
In the master integrals $J_1$-$J_3$ the logarithms $\ln(y_1)$ and $\ln(y_2)$ appear.
We would like to express them at $\bar{q}=0$ in terms of $\bar{w}_1$ and $\bar{w}_2$.
We have for $\bar{q}=0$
\bq
\label{def_boundary_L}
 L_1 
 & = & 
 \ln\left(y_1\right) 
 \; = \;
 \ln\left(\bar{w}_2\right) + 2 \ln\left(1-\bar{w}_1\right) - 2 \ln\left(1-\bar{w}_1\bar{w}_2\right),
 \nonumber \\
 L_2
 & = & 
 \ln\left(y_2\right) 
 \; = \;
 \ln\left(\bar{w}_1\right) + 2 \ln\left(1-\bar{w}_2\right) - 2 \ln\left(1-\bar{w}_1\bar{w}_2\right).
\eq
In addition, the logarithm $\ln(\Delta/\mu^4)$ will appear.
We have with $\mu=m_3$ for $\bar{q}=0$
\bq
 L_\Delta & = &
 \ln\left(\frac{\Delta}{\mu^4}\right)
 \; = \;
 \ln\left( - \frac{\left(1-\bar{w}_1\right)^2\left(1-\bar{w}_2\right)^2}{\left(1-\bar{w}_1\bar{w}_2\right)^2} \right).
\eq
Let us define the boundary constants $C_{4,j}$ through
\bq
 J_4\left(\tau=i\infty\right)
 & = &
 \sum\limits_{j=0}^\infty C_{4,j} \; \eps^j.
\eq
The sunrise integral at $\tau_0=i\infty$ is given by \cite{Adams:2014vja,Adams:2015gva}
\bq
\label{boundary_sunrise}
 J_4\left(\tau=i\infty\right)
 & = &
 \frac{1}{4}
 e^{2 \Eulerconstant \eps} \Gamma\left(1+2\eps\right)
 \left( \frac{\sqrt{\Delta}}{\mu^2} \right)^{-2\eps}
 \left[ \frac{\Gamma\left(1+\eps\right)^2}{\Gamma\left(1+2\eps\right)}
        \left( f_1+f_2+f_3\right)
        - 2 \pi \eps \right],
\eq
with
\bq
 f_j
 = 
 \frac{1}{i}
 \left[ 
  \left(-\bar{w}_j\right)^{-\eps} \; {}_2F_1\left(-2\eps,-\eps;1-\eps; \bar{w}_j \right)
  -
  \left(-\bar{w}_j^{-1}\right)^{-\eps} \; {}_2F_1\left(-2\eps,-\eps;1-\eps; \bar{w}_j^{-1} \right)
 \right].
\eq
The hypergeometric function can be expanded systematically in $\eps$ with the methods of \cite{Weinzierl:2002hv}.
The first few terms are given by
\bq
 {}_2F_1\left(-2\eps,-\eps;1-\eps; x \right)
 & = &
 1 + 2 \eps^2 \mathrm{Li}_2\left(x\right)
 + \eps^3 \left[ 2 \mathrm{Li}_3\left(x\right) - 4 \mathrm{Li}_{2,1}\left(x,1\right) \right]
 \nonumber \\
 & &
 + \eps^4 \left[ 2 \mathrm{Li}_4\left(x\right) - 4 \mathrm{Li}_{3,1}\left(x,1\right) + 8 \mathrm{Li}_{2,1,1}\left(x,1,1\right) \right]
 + {\mathcal O}\left(\eps^5\right).
 \;\;\;
\eq
The multiple polylogarithms are defined by \cite{Goncharov_no_note,Borwein,Moch:2001zr}
\bq 
\label{def_multiple_polylogs_sum}
 \mathrm{Li}_{m_1,...,m_k}(x_1,...,x_k)
  & = & \sum\limits_{n_1>n_2>\ldots>n_k>0}^\infty
     \frac{x_1^{n_1}}{{n_1}^{m_1}}\ldots \frac{x_k^{n_k}}{{n_k}^{m_k}}.
\eq
The first few boundary constants are given by
\bq
\label{def_boundary_C}
 C_{4,0} & = & 0,
 \nonumber \\
 C_{4,1} & = & 0,
 \nonumber \\
 C_{4,2} & = & 
 \sum\limits_{j=1}^3
 \frac{1}{2i} 
 \left[ \mathrm{Li}_2\left(\bar{w}_j\right) - \mathrm{Li}_2\left(\bar{w}_j^{-1}\right) \right],
 \nonumber \\
 C_{4,3} & = & 
 \sum\limits_{j=1}^3
 \frac{1}{2i} 
 \left[
  - 2 \mathrm{Li}_{2,1}\left(\bar{w}_j,1\right) - \mathrm{Li}_3\left(\bar{w}_j\right) 
  + 2  \mathrm{Li}_{2,1}\left(\bar{w}_j^{-1},1\right) + \mathrm{Li}_3\left(\bar{w}_j^{-1}\right)
 \right] 
  - L_\Delta C_{4,2},
 \nonumber \\
 C_{4,4} & = & 
 \sum\limits_{j=1}^3
 \left\{
 \frac{1}{2 i}
 \left[
    4 \mathrm{Li}_{2,1,1}\left(\bar{w}_j,1,1\right) - 2 \mathrm{Li}_{3,1}\left(\bar{w}_j,1\right) + \mathrm{Li}_4\left(\bar{w}_j\right) 
  - 4 \mathrm{Li}_{2,1,1}\left(\bar{w}_j^{-1},1,1\right) 
 \right. \right. 
 \nonumber \\
 & &
 \left. \left.
  + 2 \mathrm{Li}_{3,1}\left(\bar{w}_j^{-1},1\right) 
  - \mathrm{Li}_4\left(\bar{w}_j^{-1}\right) 
 \right]
  + \frac{\pi}{2} \left( 1 - 2 z_j \right)
        \left[ \mathrm{Li}_3\left(\bar{w}_j\right) - 2 \mathrm{Li}_{2,1}\left(\bar{w}_j,1\right) 
 \right. \right. 
 \nonumber \\
 & & 
 \left. \left.
             + \mathrm{Li}_3\left(\bar{w}_j^{-1}\right) 
             - 2  \mathrm{Li}_{2,1}\left(\bar{w}_j^{-1},1\right)
        \right]
  + i \zeta_2 \left[ 1 - 6 z_j \left(1-z_j\right)\right]
            \left[ \mathrm{Li}_2\left(\bar{w}_j\right) - \mathrm{Li}_2\left(\bar{w}_j^{-1}\right) \right]
  \right\}
 \nonumber \\
 & &
  - L_\Delta C_{4,3}
  - \frac{1}{2} L_\Delta^2 C_{4,2}
  + \pi \zeta_3.
\eq
Eq.~(\ref{result_tadpoles}) and eq.~(\ref{boundary_sunrise}) together with the condition that
the master integrals $J_4$-$J_7$ are regular at $\bar{q}=0$ fix all integration constants.
For all master integrals we write
\bq
 J_i
 & = &
 \sum\limits_{j=0}^\infty \eps^j J_i^{(j)}.
\eq
The first non-vanishing coefficients of the $\eps$-expansion of the master integrals $J_4$-$J_7$ read
\bq
 J_4^{(2)}
 & = &
 C_{4,2}
 -
  \sum\limits_{j=1}^3
 \left[
  \iterint{ \omega_0\left(\tau\right), \omega_3\left(z_j,\tau\right) }
 -4\,\iterint{ \omega_0\left(\tau\right), \omega_3\left(z_j,2\tau\right) }
 \right],
 \nonumber \\
 J_5^{(1)}
 & = &
 -L_1-L_2
 +2\,\iterint{ \omega_2\left(z_1,\tau\right) }
 +2\,\iterint{ \omega_2\left(z_2,\tau\right) }
 -4\,\iterint{ \omega_2\left(z_3,\tau\right) }
 \nonumber \\
 & &
 -4\,\iterint{ \omega_2\left(z_1,2\tau\right) }
 -4\,\iterint{ \omega_2\left(z_2,2\tau\right) }
 +8\,\iterint{ \omega_2\left(z_3,2\tau\right) },
 \nonumber \\
 J_6^{(1)}
 & = &
 -L_1+L_2
 +2\,\iterint{ \omega_2\left(z_1,\tau\right) }
 -2\,\iterint{ \omega_2\left(z_2,\tau\right) }
 -4\,\iterint{ \omega_2\left(z_1,2\tau\right) }
 +4\,\iterint{ \omega_2\left(z_2,2\tau\right) },
 \nonumber \\
 J_7^{(1)}
 & = &
 - \sum\limits_{j=1}^3 
 \left[ 
  \iterint{ \omega_3\left(z_j,\tau\right) }
 -4\,\iterint{ \omega_3\left(z_j,2\tau\right) }
 \right].
\eq
The quantities $J_i^{(j)}$ are given for $1 \le i \le 7$ and $0 \le j \le 4$
in the supplementary electronic file attached to the arxiv version of this article.

The iterated integrals in $\tau$ have a $\bar{q}$-expansion, which follows from the 
$\bar{q}$-expansion of the integration kernels by integrating term-by-term.
The $\bar{q}$-expansions of the functions $g^{(k)}(z,\tau)$ are given in eq.~(\ref{g_n_explicit}).
The $\bar{q}$-expansion of the iterated integrals provides an efficient method for the numerical evaluation
of the result \cite{Bogner:2017vim,Honemann:2018mrb}.
All results have been verified numerically with the help of the program \verb|sector_decomposition| \cite{Bogner:2007cr}.

The $\eps^0$-term of $J_4$ has been computed previously in terms of elliptic dilogarithms \cite{Adams:2014vja,Bloch:2016izu}. It is instructive to see, how the two expressions are equivalent.
One has
\bq
 J_4^{(2)}
 & = &
 C_{4,2} 
  - \sum\limits_{j=1}^3 
  \left[ \iterint{ \omega_3\left(z_j,\tau\right) }
 +4\,\iterint{ \omega_3\left(z_j,2\tau\right) } \right]
 \nonumber \\
 & = &
 C_{4,2} 
  -  
   \frac{1}{\pi} 
   \sum\limits_{j=1}^3 
   \int\limits_{i \infty}^\tau d\tau_1 \int\limits_{i \infty}^{\tau_1} d\tau_2
   \left[ g^{(3)}\left(z_j,\tau_2\right) - 8 g^{(3)}\left(z_j, 2 \tau_2 \right) \right]
 \nonumber \\
 & = &
 C_{4,2} 
 + 
 i 
 \sum\limits_{j=1}^3 
 \int\limits_0^{\bar{q}} \frac{d\bar{q}_1}{\bar{q}_1}
 \int\limits_0^{\bar{q}_1} \frac{d\bar{q}_2}{\bar{q}_2}
 \left[ \overline{\mathrm{E}}_{0;-2}\left(\bar{w}_j;1;\bar{q}_2\right) - 8 \overline{\mathrm{E}}_{0;-2}\left(\bar{w}_j;1;\bar{q}_2^2\right) \right]
 \nonumber \\
 & = &
 C_{4,2} 
 + 
 i 
 \sum\limits_{j=1}^3 
 \left[ \overline{\mathrm{E}}_{2;0}\left(\bar{w}_j;1;\bar{q}\right) - 2 \overline{\mathrm{E}}_{2;0}\left(\bar{w}_j;1;\bar{q}^2\right) \right]
 \nonumber \\
 & = &
 C_{4,2} 
 + \frac{1}{i} 
 \sum\limits_{j=1}^3 
 \overline{\mathrm{E}}_{2;0}\left(\bar{w}_j;-1;\bar{q}\right).
\eq
The expression on the last line is the result of ref.~\cite{Adams:2014vja}, taking into acount that $\bar{q}=-q_F$.

\subsection{Integration along $\tau=\mathrm{const}$}

Alternatively we may integrate the differential equation along a path with $\tau=\mathrm{const}$
from a suitable boundary point (which could be a point where one or more masses are zero).
This allows us to express the result in terms of elliptic multiple polylogarithms.
Ref.~\cite{Broedel:2017kkb} defines elliptic multiple polylogarithms 
as iterated integrals on an elliptic curve
with fixed modular parameter $\tau$:
\bq
\label{def_Gammatilde}
 \widetilde{\Gamma}\!\left({\begin{smallmatrix} n_1 & ... & n_k \\ z_1 & ... & z_k \\ \end{smallmatrix}}; z; \tau \right)
 & = &
 \int\limits_0^z dz' \; g^{(n_1)}(z'-z_1, \tau) \;
 \widetilde{\Gamma}\!\left({\begin{smallmatrix} n_2 & ... & n_k \\ z_2 & ... & z_k \\ \end{smallmatrix}}; z'; \tau \right),
\eq
where the functions $g^{(n)}(z,\tau)$ are the ones defined in eq.~(\ref{def_g_n}).
For $\tau=\mathrm{const}$ we have
\bq
 \eta_2\left(\tau\right) \;\; \stackrel{\tau=\mathrm{const}}{\longrightarrow} \;\; 0,
 \;\;\; & & \;\;\;
 \eta_4\left(\tau\right) \;\; \stackrel{\tau=\mathrm{const}}{\longrightarrow} \;\; 0,
\eq
and
\bq
 \omega_k\left(z_j,N\tau\right) & \stackrel{\tau=\mathrm{const}}{\longrightarrow} &
 \left(2\pi\right)^{2-k} g^{(k-1)}\left(z_j, N \tau\right) dz_j.
\eq
We recall that $N \in \{1,2\}$.
We would obtain immediately a solution in terms of elliptic multiple polylogarithms, if terms with $g^{(k)}(z_j, 2 \tau)$
were absent.
In order to convert the result to elliptic multiple polylogarithms we have to express
the functions $g^{(k)}(z_j, 2 \tau)$ in terms of functions $g^{(k)}(z_j', \tau)$.
From eq.~(\ref{g_n_explicit}) we find
\bq
 g^{(k)}\left(z,2\tau\right)
 & = &
 \frac{1}{2}
 \left[
 g^{(k)}\left(\frac{z}{2},\tau\right)
 +
 g^{(k)}\left(\frac{z}{2}+\frac{1}{2},\tau\right)
 \right].
\eq
After a rescaling $z'=z/2$ we obtain iterated integrals in the form of eq.~(\ref{def_Gammatilde}).


\section{Conclusions}
\label{sect:conclusions}

In this paper we considered the two-loop sunrise integral with unequal masses.
We showed that there is a basis of master integrals, in which the system of differential equations
is in $\eps$ form.
In addition we performed a change of variables for the kinematic variables, which allowed us to identify
the integration kernels as differential one-forms on the moduli space $\overline{\mathcal M}_{1,3}$.
Hence, the solution for the sunrise integral with unequal masses is given in terms of iterated
integrals on the moduli space $\overline{\mathcal M}_{1,3}$.
These iterated integrals are pure functions.

We expect our findings to have implications for a wider class of Feynman integrals.

\subsection*{Acknowledgements}

We are grateful to Luise Adams for discussions and collaboration during the initial stage of this project.

S.W. would like to thank the Institute for Theoretical Studies in Zurich
for hospitality, where part of this work was carried out.

This work has been supported by the 
Cluster of Excellence ``Precision Physics, Fundamental Interactions, and Structure of Matter'' 
(PRISMA+ EXC 2118/1) funded by the German Research Foundation (DFG) 
within the German Excellence Strategy (Project ID 39083149).


\begin{appendix}

\section{Notation for standard mathematical functions}

As notations for standard mathematical functions differ slightly in the literature, we list here the definitions and
the conventions which we follow.

The complete elliptic integral of the first kind is defined by
\bq
 K\left(k\right)
 & = &
 \int\limits_0^1 \frac{dt}{\sqrt{\left(1-t^2\right)\left(1-k^2t^2\right)}}.
\eq
The incomplete elliptic integral of the first kind is defined by
\bq
 F\left(u,k\right)
 & = &
 \int\limits_0^u \frac{dt}{\sqrt{\left(1-t^2\right)\left(1-k^2t^2\right)}}.
\eq
Dedekind's eta function is defined by
\bq
 \eta\left(\tau\right) 
 & = & 
 e^{\frac{i \pi \tau}{12}} \prod\limits_{n=1}^{\infty} \left(1-e^{2\pi i n \tau}\right).
\eq
The theta functions are defined by
\bq
\theta_1\left(z,q\right) 
 & = &
 -i \sum\limits_{n=-\infty}^\infty \left(-1\right)^n q^{\left(n+\frac{1}{2}\right)^2} e^{i\left(2n+1\right)z},
 \nonumber \\
\theta_2\left(z,q\right) 
 & = &
 \sum\limits_{n=-\infty}^\infty q^{\left(n+\frac{1}{2}\right)^2} e^{i\left(2n+1\right)z},
 \nonumber \\
\theta_3\left(z,q\right) 
 & = &
 \sum\limits_{n=-\infty}^\infty q^{n^2} e^{2 i n z},
 \nonumber \\
\theta_4\left(z,q\right) 
 & = &
 \sum\limits_{n=-\infty}^\infty \left(-1\right)^n q^{n^2} e^{2 i n z}.
\eq


\section{Details on isogenic elliptic curves}
\label{sect:details_elliptic_curves}

In this appendix we give details on the relation between 
the elliptic curve obtained from the Feynman graph polynomial
and the elliptic curve obtained from the maximal cut.

\subsection{The elliptic curve from the graph polynomial}

The second graph polynomial for the sunrise graph is given by
\bq
 {\mathcal F} & = & - \alpha_1 \alpha_2 \alpha_3 \frac{t}{\mu^2}
                + \left( \alpha_1 \frac{m_1^2}{\mu^2} + \alpha_2 \frac{m_2^2}{\mu^2} + \alpha_3 \frac{m_3^2}{\mu^2} \right) \left( \alpha_1 \alpha_2 + \alpha_2 \alpha_3 + \alpha_3 \alpha_1 \right).
\eq
The equation ${\mathcal F}=0$ defines a cubic curve in ${\mathbb P}^2({\mathbb C})$ with coordinates $[\alpha_1:\alpha_2:\alpha_3]$, and together with the choice of a rational point an elliptic curve.
Rational points are for example the three intersection points of the curve with the Feynman parameter integration region.
These points are given by
\bq
\label{intersection_F_sigma}
 P_1 = \left[1:0:0\right], 
 \;\;\;
 P_2 = \left[0:1:0\right], 
 \;\;\;
 P_3 = \left[0:0:1\right].
\eq
We will choose one of these three points $P_1$, $P_2$, $P_3$ as the origin $O$.
We denote the corresponding elliptic curves by $E_{1,F}$, $E_{2,F}$ and $E_{3,F}$, where the subscript refers to the choice of the origin:
\bq
 E_{i,F} & : & {\mathcal F} = 0 \;\;\;\mbox{with} \;\;\; O = P_i.
\eq
The elliptic curve $E_{i,F}$ (with $i \in \{1,2,3\}$) can be transformed into the Weierstrass normal form
\bq
\label{WNF_with_g2_g3}
 \hat{E}_F & : & \hat{y}^2 \hat{z} = 4 \hat{x}^3 - g_{2,F} \hat{x} \hat{z}^2 - g_{3,F} \hat{z}^3.
\eq
Under this change of variables the origin $O$ of $E_{i,F}$ is transformed to the point $[\hat{x}:\hat{y}:\hat{z}]=[0:1:0]$,
which is the origin (or the point at infinity) of the elliptic curve $\hat{E}_F$.
Note that the same Weierstrass normal form is obtained for $E_{1,F}$, $E_{2,F}$ and $E_{3,F}$.
In the following we will work in the chart $\hat{z}=1$.
Factorising the cubic polynomial on the right-hand side of eq.~(\ref{WNF_with_g2_g3}),
the Weierstrass normal form can equally be written as
\bq
 \hat{y}^2 & = & 4 \left(\hat{x}-e_{1,F}\right)\left(\hat{x}-e_{2,F}\right)\left(\hat{x}-e_{3,F}\right),
 \;\;\;\;\;\;
 \mbox{with} 
 \;\;\;
 e_{1,F}+e_{2,F}+e_{3,F}=0,
\eq
and
\bq
 g_{2,F} = -4 \left( e_{1,F} e_{2,F} + e_{2,F} e_{3,F} + e_{3,F} e_{1,F} \right),
 & &
 g_{3,F} = 4 e_{1,F} e_{2,F} e_{3,F}.
\eq
The roots are given by
\bq
\label{def_roots}
 e_{1,F} 
 & = & 
 \frac{1}{24 \mu^4} 
 \left[ -t^2 + 2 M_{100} t + \Delta 
        + 3 \left(\mu_1^2-t\right)^{\frac{1}{2}} \left(\mu_2^2-t\right)^{\frac{1}{2}} \left(\mu_3^2-t\right)^{\frac{1}{2}} \left(\mu_4^2-t\right)^{\frac{1}{2}} 
 \right],
 \nonumber \\
 e_{2,F} 
 & = & 
 \frac{1}{24 \mu^4} 
 \left[ -t^2 + 2 M_{100} t + \Delta 
        - 3 \left(\mu_1^2-t\right)^{\frac{1}{2}} \left(\mu_2^2-t\right)^{\frac{1}{2}} \left(\mu_3^2-t\right)^{\frac{1}{2}} \left(\mu_4^2-t\right)^{\frac{1}{2}}
 \right],
 \nonumber \\
 e_{3,F} 
 & = & 
 \frac{1}{24 \mu^4} \left( 2 t^2 - 4 M_{100} t - 2 \Delta \right).
\eq
We set
\bq
 Z_{1,F} \; = \; e_{3,F} - e_{2,F},
 \;\;\;\;\;\;
 Z_{2,F} \; = \; e_{1,F} - e_{3,F},
 \;\;\;\;\;\;
 Z_{3,F} \; = \; e_{1,F} - e_{2,F},
\eq
and 
\bq
 k_F^2 \; = \; \frac{Z_{1,F}}{Z_{3,F}},
 & &
 \bar{k}_F^2 \; = \; \frac{Z_{2,F}}{Z_{3,F}}.
\eq
We define two periods
\bq
\label{def_periods}
 \psi_{1,F} =  
 2 \int\limits_{e_{2,F}}^{e_{3,F}} \frac{dx}{y}
 =
 \frac{2}{Z_{3,F}^{\frac{1}{2}}} K\left(k_F\right),
 & &
 \psi_{2,F} =  
 2 \int\limits_{e_{1,F}}^{e_{3,F}} \frac{dx}{y}
 =
 \frac{2 i}{Z_{3,F}^{\frac{1}{2}}} K\left(\bar{k}_F\right),
\eq
together with the modular parameter $\tau_F$, the nome $q_F$ and the nome squared $\bar{q}_F$:
\bq
 \tau_F \; = \; \frac{\psi_{2,F}}{\psi_{1,F}},
 & &
 q_F \; = \; e^{\pi i \tau_F},
 \;\;\;\;\;\;
 \bar{q}_F \; = \; e^{2 \pi i \tau_F}.
\eq
The lattice $\Lambda_F$ is given by
\bq
 \Lambda_F & = & \{ n_1  + n_2 \tau_F \; | \; n_1, n_2 \in {\mathbb Z} \}.
\eq
In the Euclidean region, defined by $t<0$ and $m_1^2, m_2^2, m_3^2 > 0$,
the period $\psi_{1,F}$ is real and the period $\psi_{2,F}$ purely imaginary.
In the Euclidean region $\Lambda_F$ is a rectangular lattice.

\subsection{The elliptic curve from the maximal cut}

For the maximal cut we consider the Baikov representation \cite{Baikov:1996iu}, 
and here in particular the loop-by-loop approach \cite{Frellesvig:2017aai}.
Within the loop-by-loop approach there are for the sunrise integral
three possibilities to choose the first loop, obtained by choosing two out of the three propagators.
The second loop contains then necessarily the propagator not selected for the first loop.
We thus obtain three elliptic curves
\bq
 E_{1,C} 
 & : &
 v^2 = 
 \left[ u + \frac{\left(m_2+m_3\right)^2}{\mu^2}\right]
 \left[ u + \frac{\left(m_2-m_3\right)^2}{\mu^2}\right]
 \left[ u^2 + 2 \frac{\left(t+m_1^2\right)}{\mu^2} u + \frac{\left(t-m_1^2\right)^2}{\mu^4}\right],
 \nonumber \\
 E_{2,C} 
 & : &
 v^2 = 
 \left[ u + \frac{\left(m_1+m_3\right)^2}{\mu^2}\right]
 \left[ u + \frac{\left(m_1-m_3\right)^2}{\mu^2}\right]
 \left[ u^2 + 2 \frac{\left(t+m_2^2\right)}{\mu^2} u + \frac{\left(t-m_2^2\right)^2}{\mu^4}\right],
 \nonumber \\
 E_{3,C} 
 & : &
 v^2 = 
 \left[ u + \frac{\left(m_1+m_2\right)^2}{\mu^2}\right]
 \left[ u + \frac{\left(m_1-m_2\right)^2}{\mu^2}\right]
 \left[ u^2 + 2 \frac{\left(t+m_3^2\right)}{\mu^2} u + \frac{\left(t-m_3^2\right)^2}{\mu^4}\right],
\eq
which differ by a permutation of the particle masses.
Let us consider $E_{3,C}$, which corresponds to the maximal cut given in eq.~(\ref{maxcut}).
We denote the roots of the quartic polynomial by
\bq
 u_1 = - \frac{\left(m_1+m_2\right)^2}{\mu^2},
 \;\;\;
 u_2 = - \frac{\left(m_3 + \sqrt{t}\right)^2}{\mu^2},
 \;\;\;
 u_3 = - \frac{\left(m_3 - \sqrt{t}\right)^2}{\mu^2},
 \;\;\;
 u_4 = - \frac{\left(m_1-m_2\right)^2}{\mu^2}.
\eq
We set
\bq
 Z_{1,C} \; = \; \left(u_3-u_2\right)\left(u_4-u_1\right),
 \;\;\;\;\;\;
 Z_{2,C} \; = \; \left(u_2-u_1\right)\left(u_4-u_3\right),
 \;\;\;\;\;\;
 Z_{3,C} \; = \; \left(u_3-u_1\right)\left(u_4-u_2\right),
 \;\;\;
\eq
and 
\bq
\label{def_k_C}
 k_C^2 \; = \; \frac{Z_{1,C}}{Z_{3,C}},
 & &
 \bar{k}_C^2 \; = \; \frac{Z_{2,C}}{Z_{3,C}}.
\eq
We define two periods
\bq
\label{def_periods_C}
 \psi_{1,C} =  2 \int\limits_{u_2}^{u_3} \frac{du}{v}
 =
 \frac{4}{Z_{3,C}^{\frac{1}{2}}} K\left(k_C\right),
 & &
 \psi_{2,C} =  2 \int\limits_{u_4}^{u_3} \frac{du}{v}
 =
 \frac{4 i}{Z_{3,C}^{\frac{1}{2}}} K\left(\bar{k}_C\right),
\eq
together with the modular parameter $\tau_C$, the nome $q_C$ and the nome squared $\bar{q}_C$:
\bq
\label{def_tau_C}
 \tau_C \; = \; \frac{\psi_{2,C}}{\psi_{1,C}},
 & &
 q_C \; = \; e^{\pi i \tau_C},
 \;\;\;\;\;\;
 \bar{q}_C \; = \; e^{2 \pi i \tau_C}.
\eq
The lattice $\Lambda_C$ is given by
\bq
 \Lambda_C & = & \{ n_1  + n_2 \tau_C \; | \; n_1, n_2 \in {\mathbb Z} \}.
\eq
We may transform $E_{3,C}$ into the Weierstrass normal form
\bq
 \hat{E}_C & : & \hat{y}^2 \hat{z} \; = \; 4 \hat{x}^3 - g_{2,C} \hat{x} \hat{z}^2 - g_{3,C} \hat{z}^3.
\eq
We use the chart $\hat{z}=1$ and factor the cubic polynomial
\bq
 \hat{y}^2 & = & 4 \left(\hat{x}-e_{1,C}\right)\left(\hat{x}-e_{2,C}\right)\left(\hat{x}-e_{3,C}\right).
\eq
$g_{2,C}$ and $g_{3,C}$ are given by
\bq
 g_{2,C} = -4 \left( e_{1,C} e_{2,C} + e_{2,C} e_{3,C} + e_{3,C} e_{1,C} \right),
 & &
 g_{3,C} = 4 e_{1,C} e_{2,C} e_{3,C}.
\eq
The roots are given by
\bq
 e_{1,C} 
 & = & 
 \frac{1}{3\mu^4} \left(  t^2 - 2 M_{100} t - \Delta + 24 m_1 m_2 m_3 \sqrt{t} \right),
 \nonumber \\
 e_{2,C} 
 & = & 
 \frac{1}{3\mu^4} \left(  t^2 - 2 M_{100} t - \Delta - 24 m_1 m_2 m_3 \sqrt{t} \right),
 \nonumber \\
 e_{3,C} 
 & = & 
 \frac{1}{3\mu^4} \left(  - 2 t^2 + 4 M_{100} t + 2 \Delta \right).
\eq
We have
\bq
 Z_{1,C} \; = \; e_{1,C} - e_{2,C},
 \;\;\;\;\;\;
 Z_{2,C} \; = \; e_{3,C} - e_{1,C},
 \;\;\;\;\;\;
 Z_{3,C} \; = \; e_{3,C} - e_{2,C}.
\eq

\subsection{The relation between $\Lambda_F$ and $\Lambda_C$}

\begin{figure}
\begin{center}
\includegraphics[scale=1.0]{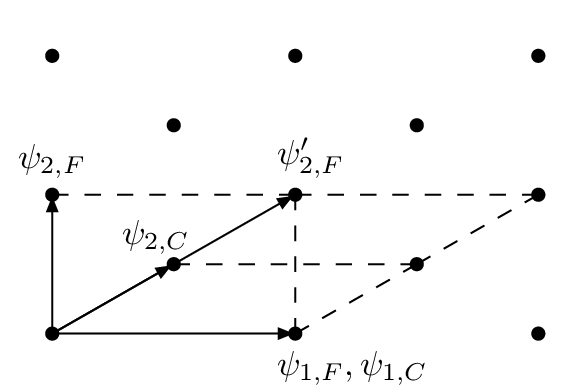}
\end{center}
\caption{
The periods $(\psi_{1,C},\psi_{2,C})$ generate a lattice.
The periods $(\psi_{1,F},\psi_{2,F})$ or $(\psi_{1,F},\psi_{2,F}')$ generate a sub-lattice of index $2$.
Fundamental cells of the various lattices are indicated by dashed lines.
}
\label{fig_lattice}
\end{figure}

We may now compare the lattices $\Lambda_F$ and $\Lambda_C$.
We find
\bq
 \psi_{1,C} \; = \; \psi_{1,F},
 \;\;\;\;\;\;
 2 \psi_{2,C} \; = \; \psi_{2,F} + \psi_{1,F}.
\eq
This shows that $\Lambda_F$ is a sub-lattice of $\Lambda_C$ of index $2$.
This becomes apparent if we define
\bq
 \left( \begin{array}{c}
 \psi_{2,F}' \\
 \psi_{1,F}' \\
 \end{array} \right)
 & = &
 \left( \begin{array}{cc}
 1 & 1 \\
 0 & 1 \\
 \end{array} \right)
 \left( \begin{array}{c}
 \psi_{2,F} \\
 \psi_{1,F} \\
 \end{array} \right).
\eq
We have
\bq
\label{def_periods_F_prime}
 \psi_{1,F}'
 \; = \;
 \frac{2}{\sqrt{Z_{2,F}}}
 K\left(\sqrt{\frac{-Z_{1,F}}{Z_{2,F}}}\right),
 & &
 \psi_{2,F}'
 \; = \;
 \frac{2i}{\sqrt{Z_{2,F}}}
 K\left(\sqrt{\frac{Z_{3,F}}{Z_{2,F}}}\right),
\eq
and
\bq
 \psi_{1,C}
 \; = \;
 \psi_{1,F}',
 & &
 2 \psi_{2,C}
 \; = \;
 \psi_{2,F}'.
\eq
The geometric situation is shown in fig.~\ref{fig_lattice}.
With
\bq
 \tau_F' \; = \; \frac{\psi_{2,F}'}{\psi_{1,F}'},
 & &
 q_F' \; = \; e^{\pi i \tau_F'},
 \;\;\;\;\;\;
 \bar{q}_F' \; = \; e^{2\pi i \tau_F'},
\eq
we further have
\bq
 2 \tau_C \; = \; \tau_F + 1 \; = \; \tau_F',
 & &
 q_C^2 \; = \; - q_F \; = \; q_F',
 \;\;\;\;\;\;
 \bar{q}_C^2 \; = \; \bar{q}_F \; = \; \bar{q}_F'.
\eq

\subsection{Marked points}
\label{subsec:marked_points}

On the curves $E_{i,F}$ we have three rational points $P_1$, $P_2$ and $P_3$ given by the intersection of the curve with 
the integration region in Feynman parameter space.
These determine three points on $\hat{E}_F$ and in ${\mathbb C}/\Lambda_F$.
In this paragraph we determine the corresponding points in ${\mathbb C}/\Lambda_C$ and on $\hat{E}_C$.
This is done as follows:
Starting from the elliptic curve $E_{3,F}$, which is given as the cubic curve ${\mathcal F}=0$ together
with the choice $O=P_3$ as origin, we can go over to the elliptic curve $\hat{E}_F$ in Weierstrass normal form.
From there we can move on to a torus ${\mathbb C}/\Lambda_F$, where $\Lambda_F$ is the lattice defined by the
periods of the elliptic curve $\hat{E}_F$.
There is map from ${\mathbb C}/\Lambda_F$ to ${\mathbb C}/\Lambda_C$, sending $z_F \in {\mathbb C}/\Lambda_F$ 
to $z_C = (z_F \mod \Lambda_C ) \in {\mathbb C}/\Lambda_C$.
From ${\mathbb C}/\Lambda_C$ we map back to $\hat{E}_C$.
In principle we can further map back to $E_{3,C}$, but we will not need this.
We thus have a sequence of mappings
\bq
\label{chain_of_mappings}
\begin{CD}
 E_{3,F} @>>> \hat{E}_F @>>> {\mathbb C}/\Lambda_F \\
 & & & & @VVV \\
 E_{3,C} @<<< \hat{E}_C @<<< {\mathbb C}/\Lambda_C
\end{CD}
\eq
On the Feynman graph polynomial side it is easier to work with the periods $\psi_{1,F}'$ and $\psi_{2,F}'$,
defined in eq.~(\ref{def_periods_F_prime}).
We set
\bq
 k_F'{}^2 & = & \frac{-Z_{1,F}}{Z_{2,F}}.
\eq
The elliptic curve $E_{3,F}$ is defined by the cubic curve ${\mathcal F}=0$ and the choice $O=P_3$ as origin.
We denote by $Q_{i,j,F}$ the image of the point $P_i \in E_{j,F}$ on $\hat{E}_F$.
By construction we have
\bq
 Q_{i,i,F} & = & \left[ 0 : 1 : 0 \right].
\eq
The points $Q_{1,3,F}$ and $Q_{2,3,F}$ are given by
\bq
\label{points_on_E_hat}
 Q_{1,3,F} & = & 
 \left[e_{3,F}+\frac{m_1^2 m_3^2}{\mu^4}:-\frac{m_1^2 m_3^2\left(t-m_1^2+m_2^2-m_3^2\right)}{\mu^6}:1\right],
 \nonumber \\
 Q_{2,3,F} & = & 
 \left[e_{3,F}+\frac{m_2^2 m_3^2}{\mu^4}:\frac{m_2^2 m_3^2\left(t+m_1^2-m_2^2-m_3^2\right)}{\mu^6}:1\right].
\eq
We denote the coordinates of $Q_{i,j,F}$ by $[\hat{x}_{i,j,F} : \hat{y}_{i,j,F} : 1 ]$. 
We have
\bq
 \hat{x}_{i,j,F} = \hat{x}_{j,i,F},
 & &
 \hat{y}_{i,j,F} = - \hat{y}_{j,i,F},
\eq
and therefore we may write
\bq
 Q_{i,j,F} & = & - Q_{j,i,F}
\eq
with respect to the addition on $\hat{E}_F$.
In particular we have
\bq
 Q_{3,1,F} & = & 
 \left[e_{3,F}+\frac{m_1^2 m_3^2}{\mu^4}:\frac{m_1^2 m_3^2\left(t-m_1^2+m_2^2-m_3^2\right)}{\mu^6}:1\right].
\eq
The mapping from $\hat{E}_F$ to ${\mathbb C}/\Lambda_F$ is given by
\bq
 \left[\hat{x}_F:\hat{y}_F:1\right] 
 & \rightarrow &
 z_F =
 \frac{1}{\psi_{1,F}'} 
 \int\limits_{\hat{x}_F}^\infty 
 \frac{d\tilde{x}}{\sqrt{4\left(\tilde{x}-e_{1,F}\right)\left(\tilde{x}-e_{2,F}\right)\left(\tilde{x}-e_{3,F}\right)}}.
\eq
The integral is an incomplete elliptic integral of the first kind.
Transforming this integral into the standard form, we find 
that the points $Q_{1,2,F}$, $Q_{2,3,F}$ and $Q_{3,1,F}$ are mapped to
\bq
\label{def_coordinate_torus}
 Q_{j,k,F}
 & \rightarrow &
 z_{i,F} =
 \frac{F\left(u_{i,F}',k_F'\right)}{2 K\left(k_F'\right)},
 \;\;\;\;\;\;
 u_{i,F}' = \sqrt{\frac{e_{1,F}-e_{3,F}}{\hat{x}_{j,k,F}-e_{3,F}}}.
\eq
Here we used the convention that the triple $(i,j,k)$ is a cyclic permutation of $(1,2,3)$.
The function $F(u,k)$ denotes the incomplete elliptic integral of the first kind.
The points $Q_{2,1,F}$, $Q_{3,2,F}$ and $Q_{1,3,F}$ are mapped to
\bq
 Q_{k,j,F}
 & \rightarrow &
 - \hat{z}_i.
\eq
Let us now consider $\Lambda_C$. We have
\bq
 2 \Lambda_C & \subset & \Lambda_F.
\eq
We identify the points $z_{i,C}$ on ${\mathbb C}/\Lambda_C$ with corresponding points $z_{i,F}$ on ${\mathbb C}/\Lambda_F$:
\bq
 z_{i,C} & = & z_{i,F} \mod \Lambda_C,
 \;\;\;\;\;\;\;\;\;
 i \in \{1,2,3\}.
\eq
We now construct points
\bq
 Q_{j,k,C} & = & \left[ \hat{x}_{j,k,C} : \hat{y}_{j,k,C} : 1 \right] \in \hat{E}_C,
\eq
such that
\bq
\label{def_coordinate_torus_C}
 Q_{j,k,C}
 & \rightarrow &
 z_{i,C} =
 \frac{F\left(u_{i,C},k_C\right)}{2 K\left(k_C\right)},
 \;\;\;\;\;\;
 u_{i,C} = \sqrt{\frac{e_{3,C}-e_{2,C}}{\hat{x}_{j,k,C}-e_{2,C}}}.
\eq
We recall that we assume that $(i,j,k)$ is a cyclic permutation of $(1,2,3)$.
We do this in two steps:
We first use a Landen transformation to find coordinates $[ \hat{x}_{j,k,C,1/2} : \hat{y}_{j,k,C,1/2} : 1 ] \in \hat{E}_C$
for $z_{i,C}/2 \in {\mathbb C}/\Lambda_C$.
In step two we use multiplication by $2$ on $\hat{E}_C$.
We start with an auxiliary map from ${\mathbb C}/\Lambda_C$ to ${\mathbb C}/\Lambda_F$, defined by
\bq
 z_{F} & = & 2 z_{C}.
\eq
In the equal mass case ($m_1=m_2=m_3$) we have
\bq
 z_{1,F} \; = \; z_{2,F} \; = \; z_{3,F} \; = \; \frac{1}{3}.
\eq
In a neighbourhood of the equal mass point we may invert the map
from ${\mathbb C}/\Lambda_C$ to ${\mathbb C}/\Lambda_F$ and we obtain
\bq
 z_{i,C,1/2} & = & \frac{z_{i,F}}{2},
 \;\;\;\;\;\;\;\;\;
 i \in \{1,2,3\}.
\eq
Similar to eq.~(\ref{def_coordinate_torus}) and eq.~(\ref{def_coordinate_torus_C}) we write
\bq
 z_{i,C,1/2}
 & = &
 \frac{F\left(u_{i,C,1/2},k_C\right)}{2 K\left(k_C\right)}.
\eq
We obtain $u_{i,C,1/2}$ as follows:
We first note that the relation between $k_F'$ and $k_C$ is given by
\bq
 k_C 
 & = &
 \frac{2\sqrt{k_F'}}{1+k_F'}.
\eq
Thus we may use the Landen transformation
\bq
 F\left(u,k\right) 
 \; = \;
 \frac{2}{1+k} F\left(u',k'\right),
 & &
 k' \; = \; \frac{2\sqrt{k}}{1+k},
 \nonumber \\
 & &
 u' \; = \; \sqrt{ \frac{1}{2} \left( 1 + k u^2- \sqrt{ 1 -\left(1+k^2\right)u^2 + k^2 u^4}\right)},
\eq
to relate $u_{i,F}'$ to $u_{i,C,1/2}$.
We find
\bq
\label{def_u_C}
 u_{1,C,1/2}
 & = &
 \sqrt{\frac{\left(\sqrt{t}+m_1-m_2+m_3\right)\left(\sqrt{t}+m_1+m_2-m_3\right)}{4 m_2 m_3}},
 \nonumber \\
 u_{2,C,1/2}
 & = &
 \sqrt{\frac{\left(\sqrt{t}-m_1+m_2+m_3\right)\left(\sqrt{t}+m_1+m_2-m_3\right)}{4 m_1 m_3}}.
\eq
From 
\bq
 u_{i,C,1/2} & = & \sqrt{\frac{e_{3,C}-e_{2,C}}{\hat{x}_{j,k,C,1/2}-e_{2,C}}}
\eq
we obtain
\bq
 \hat{x}_{2,3,C,1/2}
 & = &
 e_{2,C} - \frac{4 m_2 m_3 \left(\sqrt{t}-m_1+m_2+m_3\right)\left(\sqrt{t}-m_1-m_2-m_3\right)}{\mu^4},
 \nonumber \\
 \hat{x}_{3,1,C,1/2}
 & = &
 e_{2,C} - \frac{4 m_1 m_3 \left(\sqrt{t}+m_1-m_2+m_3\right)\left(\sqrt{t}-m_1-m_2-m_3\right)}{\mu^4}.
\eq
In the second step we construct from the point
\bq
 Q_{j,k,C,1/2} & = & \left[ \hat{x}_{j,k,C,1/2} : \hat{y}_{j,k,C,1/2} : 1 \right]
\eq
the point
\bq
 2 Q_{j,k,C,1/2} & = & Q_{j,k,C} \; = \; \left[ \hat{x}_{j,k,C} : \hat{y}_{j,k,C} : 1 \right].
\eq
For a point $P=[x:y:1]$ on the curve $y^2-4x^3+g_2x+g_3$ the point $2P$ is given by
\bq
 2 P
 & = &
 \left[
 \frac{1}{4} \left( \frac{12 x^2 - g_2}{2 y} \right)^2 - 2 x
 :
 - \frac{1}{4} \left( \frac{12 x^2 - g_2}{2 y} \right)^3
 + 3 x \left( \frac{12 x^2 - g_2}{2 y} \right) 
 - y
 : 
 1
 \right].
\eq
Thus
\bq
 \hat{x}_{2,3,C}
 & = &
 e_{2,C}
 + \frac{4 \left(m_1 \sqrt{t} + m_2 m_3 \right)^2}{\mu^4},
 \nonumber \\
 \hat{x}_{3,1,C}
 & = &
 e_{2,C}
 + \frac{4 \left(m_2 \sqrt{t} + m_1 m_3 \right)^2}{\mu^4},
\eq
and
\bq
 u_{1,C}
 & = &
 \frac{\sqrt{\left(\mu_1+\sqrt{t}\right)\left(\mu_2+\sqrt{t}\right)\left(\mu_3+\sqrt{t}\right)\left(\mu_4-\sqrt{t}\right)}}{2\left( m_1 \sqrt{t} + m_2 m_3 \right)},
 \nonumber \\
 u_{2,C}
 & = &
 \frac{\sqrt{\left(\mu_1+\sqrt{t}\right)\left(\mu_2+\sqrt{t}\right)\left(\mu_3+\sqrt{t}\right)\left(\mu_4-\sqrt{t}\right)}}{2\left( m_2 \sqrt{t} + m_1 m_3 \right)}.
\eq


\section{Supplementary material}
\label{sect:supplement}

Attached to the arxiv version of this article is an electronic file in ASCII format with {\tt Maple} syntax, defining the quantities
\begin{center}
 \verb|A|, \; \verb|J|.
\end{center}
The matrix \verb|A| appears in the differential equation
\bq
 d \vec{J} & = & \eps A \vec{J}.
\eq
The entries of the matrix $A$ are linear combinations of 
\bq
 \omega_k\left(z_j, N \tau \right),
 & &
 0 \le k \le 4, 
 \;\;\;
 1 \le j \le 3,
 \;\;\;
 1 \le N \le 2,
\eq
and $\eta_2(\tau)$, $\eta_4(\tau)$.
The vector \verb|J| contains the results for the master integrals up to order $\eps^4$ in terms of iterated integrals
in the variable $\tau$ as discussed in section~\ref{subsect:tau_integration}.
The variable $\eps$ is denoted by \verb|eps|, $\zeta_2$, $\zeta_3$, $\zeta_4$ by
\begin{center}
 \verb|zeta_2|, \verb|zeta_3|, \verb|zeta_4|,
\end{center}
respectively. \verb|L1|, \verb|L2|, \verb|C_4_2|, \verb|C_4_3| and \verb|C_4_4| denote the boundary values
defined in eq.~(\ref{def_boundary_L}) and eq.~(\ref{def_boundary_C}).

\end{appendix}

{\footnotesize
\bibliography{/home/stefanw/notes/biblio}
\bibliographystyle{/home/stefanw/latex-style/h-physrev5}
}

\end{document}